\documentclass[journal]{IEEEtran}

\usepackage[dvips]{pstcol}
\usepackage{graphicx}
\usepackage{amsmath}

\usepackage{latexsym}
\usepackage{amsfonts}
\usepackage{amsbsy}
\usepackage{amssymb}
\usepackage{times}
\usepackage{enumerate}

\usepackage{subeqnarray}
\usepackage{cases}
\usepackage{stfloats}
\usepackage{cases}
\usepackage{subeqnarray}
\usepackage{amsmath}
\usepackage{multicol}
\usepackage{multirow}
\usepackage{bigstrut}
\usepackage{cite}
\usepackage{epstopdf}

\begin{document}
\title{Joint Transceiver Design Algorithms for Multiuser MISO Relay Systems with Energy Harvesting}

%
%
%
%
%
%
%

\author{Yunlong Cai, Ming-Min Zhao, Qingjiang Shi,  Benoit Champagne and Min-Jian Zhao
\thanks{
Y. Cai, M. M. Zhao and M. J. Zhao are with the College of Information Science and Electronic Engineering, Zhejiang University, Hangzhou 310027, China (e-mail: ylcai@zju.edu.cn; zmmblack@zju.edu.cn; mjzhao@zju.edu.cn).

Q. Shi is with the School of Info. Sci. \& Tech., Zhejiang Sci-Tech University, Hangzhou 310018, China. He is also with The State Key Laboratory of Integrated Services Networks, Xidian University (email: qing.j.shi@gmail.com).

B. Champagne is with the Department of Electrical and Computer Engineering, McGill University, Montreal, QC H3A 0E9, Canada (benoit.champagne@mcgill.ca).

This work was supported in part by the National Natural Science Foundation of China under Grant  61471319, Zhejiang Provincial Natural Science Foundation of China under Grant LY14F010013, the Fundamental Research Funds for the Central Universities, and the National High Technology Research and Development Program (863 Program) of China under Grant 2014AA01A707.
}
}

\maketitle
\begin{abstract}
In this paper, we investigate a multiuser multiple-input single-output (MISO) relay system with simultaneous wireless information and power transfer (SWIPT), where the received signal is divided into two parts for information decoding (ID) and energy harvesting (EH), respectively. Assuming that both base station (BS) and relay station (RS) are equipped with multiple antennas, this work studies the joint transceiver design problem for the BS beamforming vectors, the RS amplify-and-forward (AF) transformation matrix and the power splitting (PS) ratios at the single-antenna receivers. The aim is to minimize the total transmission power of the BS and the RS under both signal-to-interference-plus-noise ratio (SINR) and EH constraints. Firstly, an iterative algorithm based on alternating optimization (AO) and with guaranteed convergence is proposed to successively optimize the transceiver coefficients. This AO-based approach is then extended into a robust transceiver design against quantization errors in channel state information (CSI), by using semidefinite relaxation (SDR) and the S-procedure. Secondly, a novel design scheme based on switched relaying (SR) is proposed that can significantly reduce the computational complexity and overhead of the AO-based designs while maintaining a similar performance. In the proposed SR scheme, the RS is equipped with a codebook of permutation matrices. For each permutation matrix, a latent transceiver is designed which consists of BS beamforming vectors, optimally scaled RS permutation matrix and receiver PS ratios. For the given CSI, the optimal transceiver with the lowest total power consumption is selected for transmission.
We propose a concave-convex procedure (CCCP) based and subgradient-type iterative algorithms for the non-robust and robust latent transceiver designs.
Simulation results are presented to validate the effectiveness of all the proposed algorithms.
\end{abstract}


\IEEEpeerreviewmaketitle

\section{Introduction}
Recently, electromagnetic (EM) energy transfer techniques have attracted considerable interest in the wireless research community. Indeed, by exploiting the radiative far-field properties of EM waves, these techniques could in theory enable a radio receiver to harvest energy from its environment, thereby relaxing the battery requirements on user devices. An interesting application of wireless energy transfer is to jointly transmit information and energy using the same waveform, which is known as simultaneous wireless information and power transfer (SWIPT). The idea of SWIPT was first proposed by Varshney in \cite{varshney2008transporting}, which characterizes the rate-energy (R-E) tradeoff in a discrete memoryless channel. The study of the R-E tradeoff was later extended to frequency selective fading channels in \cite{grover2010shannon}. In \cite{Twowaycommunication}, the authors studied a two-way communication scenario with noiseless channels and limited resources, with emphasis on the tradeoffs due to the need of balancing the information flow with the resulting energy exchange among the communicating nodes. The authors of \cite{OntheTransferofInformation} focused on multiuser systems and demonstrated that energy transfer constraints call for additional coordination among distributed nodes of a wireless network. However, the above studies have not been realized yet due to practical circuit limitations.

The first practical receiver structure that makes SWIPT possible was proposed in \cite{MIMOBroadcastingfor}, where two practical signal separation schemes were considered, namely, the time switching (TS) scheme, where the receiver switches between information decoding (ID) and energy harvesting (EH), and the power splitting (PS) scheme, where the received signal is split into two streams, such that a fraction $\rho$ ($0 \leq \rho \leq 1)$ of the received signal power is used for ID while the remaining fraction $(1-\rho)$ is used for EH.
PS-based algorithms were considered in \cite{Liang2013Wireless, Zhou2013Wireless, Shi2014Joint, BeamformingforMISO, JointBeamformingAnd}. In \cite{Liang2013Wireless}, the authors derived the optimal PS rule at the receiver, for both SISO (single-input single-output) and SIMO (single-input multiple-output) systems, in order to optimize the R-E performance tradeoff. In \cite{Zhou2013Wireless}, the authors proposed two practical receiver architectures, i.e., separated and integrated information and energy receivers, based on dynamic power splitting. These authors characterized the R-E performance by taking circuit power consumption into account.

The authors of \cite{Shi2014Joint} studied the joint beamforming and power splitting (JBPS) design for a multiuser multiple-input single-output (MISO) broadcast system with SWIPT. In this study, the total transmission power at the base station (BS) is minimized subject to signal-to-interference-plus-noise ratio (SINR) and EH constraints for all the receivers. The JBPS problem for a $K$-user MISO interference channel was considered in \cite{BeamformingforMISO}, where the authors used the semidefinite relaxation (SDR) technique to address the non-convex problem and proved that the SDR is tight in the case of $K=2$ or $3$. Different from the approach of \cite{BeamformingforMISO},
an alternative second-order-cone-programming (SOCP) relaxation method was proposed in \cite{JointBeamformingAnd}.
The SOCP relaxation based method guarantees a feasible solution to the JBPS problem and has lower complexity than the SDR method, while achieving a performance extremely close to the minimum transmission power. A primal-decomposition based decentralized algorithm was also presented in this work.

Besides, the SWIPT technique for relay systems was considered in \cite{EnergyharvestinginanOSTBC, Li2014Beamforming, Guobing2014High, Chen2015Distributed}. Specifically, the authors of \cite{EnergyharvestinginanOSTBC} proposed a joint source and relay precoding design algorithm to achieve different tradeoffs between the energy transfer and the information rate. In \cite{Li2014Beamforming}, the relay beamforming design problem for the SWIPT scheme was considered in a non-regenerative two-way multi-antenna relay network, where a global optimal solution, a local optimal solution and a low-complexity suboptimal solution were proposed. The design of a high-rate beamformer that supports multiple communication pairs and intended for SWIPT in wireless relay networks was considered in \cite{Guobing2014High}. In \cite{Chen2015Distributed}, a game-theoretical framework was developed to address the distributed power splitting problem for SWIPT in relay interference channels. The aforementioned works assume the availability of perfect channel state information (CSI), which may be impractical in realistic implementations of these systems.

In this paper, we consider the use of PS-based receivers in a general multiuser MISO relay systems, i.e., where a multi-antenna relay station (RS) is incorporated into the traditional MISO
multiuser setup, as in \cite{Tang2006Onachievable,Gomadam2007Ontheduality, zhang2009joint}. We focus our study on the downlink transmission where the BS first transmits the signals intended for different receivers via beamforming to the RS. Then, the RS processes the received signals through an amplify-and-forward (AF) transformation matrix and broadcasts it to all the receivers. Finally, each receiver employs the PS technique to decode information and harvest energy simultaneously. We present an optimization framework for the joint design of the BS beamforming vectors, the RS AF transformation matrix and the receiver PS ratios aiming to minimize the total power consumption under a set of minimum SINR and EH constraints at the receivers. In this work, we shall assume that the proposed joint design algorithms are implemented at the BS.\footnote{In practice, in cellular systems it is preferable to implement most of the signal processing operations at the BS rather than the RS due to the fact that the BS is more powerful and the RS is expected to have a simple structure and low energy consumption \cite{Dfeng1,gyli,mBanerjee}.}

To this end, we first propose an iterative algorithm based on alternating optimization (AO) and with guaranteed convergence to successively optimize the transceiver coefficients, i.e. the BS beamformers, the AF transformation matrix and the receiver PS ratios. We show that each subproblem can be relaxed as a semidefinite programming (SDP) problem by applying the celebrated SDR technique \cite{Semidefiniterelaxation}.
This AO-based approach is then extended into a robust transceiver design against quantization errors in CSI based on the SDR technique and the S-procedure \cite{ConvexOptimization, boyd1994linear}. While the performance of the AO-based designs is remarkable, we note that in this approach, the BS needs to compute and send the complete AF matrix to the RS before transmission, which entails high design complexity and signaling requirements for the overhead. Secondly, to reduce the design and implementation costs, we present extensions of the switched relaying (SR) processing reported in \cite{Cai2014TVT} to multiuser MISO relay systems with energy harvesting.
For this scheme, we equip the RS with a codebook of permutation matrices.
Based on each element of the codebook, we can obtain a permuted channel matrix and therefore create a latent transceiver, which includes
the BS beamforming vectors, an optimally scaled relaying permutation matrix and the receiver PS ratios. Among the latent transceivers so obtained for given CSI, the optimal one is chosen according to a suitable criterion for transmission.

Specifically, SR-based algorithm constructs the RS AF transformation matrix within each latent transceiver by multiplying a permutation matrix from the codebook with a power scaling factor. Before data transmission, the BS sends the index of the permutation matrix (instead of the RS AF transformation matrix), the RS power scaling factor, and the PS ratios corresponding to the optimal transceiver to the RS and the receivers through signaling channels. Compared to the AO-based algorithms, the SR approach reduces the number of optimization variables from the complete RS AF transformation matrix to a single power scaling factor, which in turn significantly reduces the computational complexity and signaling overhead.
To design the latent transceivers in the SR scheme, an iterative algorithm based on the concave-convex procedure (CCCP) \cite{yuille2003concave, lanckriet2009convergence} is proposed that uses the estimated CSI, and is guaranteed to converge to a local optimal point. Each subproblem in the iterative algorithm can be formulated as an SOCP problem. By taking the quantization CSI errors into account, we propose a robust subgradient algorithm which utilizes the side information provided by standard convex solvers. Furthermore, a simplified SR-based transceiver design algorithm is proposed with much less computational complexity, while achieving a performance extremely close to the non-simplified scheme. Finally, two efficient codebook design approaches are developed.

Simulation results demonstrate that the proposed transceiver design algorithms are capable of providing robustness against the effects of norm-bounded CSI errors. In particular, the SR-based algorithms using a small codebook of permutation matrices can achieve almost the same performance as the AO-based algorithms but with reduced design/implementation complexity and signaling overhead.

The reminder of this paper is organized as follows. Section \ref{syste-mmodel} presents the multiuser MISO relay system model, the channel error model and the problem formulation. In Section \ref{AO}, the AO-based transceiver design algorithms for both non-robust and robust cases are developed. Section \ref{SR} discusses the SR-based transceiver scheme, including the proposed latent transceiver design algorithms and the codebook design methods. A complexity analysis of the proposed algorithms along with a discussion of their initialization are provided in Section \ref{sec-complexity}. Finally, in Section \ref{Simulation} computer simulations are used to verify the proposed algorithms. Conclusions are drawn in Section \ref{conclusion}.

\emph{Notations:} Scalars, vectors and matrices are respectively denoted by lower case, boldface lower case and boldface upper case letters. For a square matrix $\bf{A}$, $\textrm{Tr} (\bf{A})$, $\textrm{rank}\bf(A)$, ${{\bf{A}}^T}$, $\textrm{conj}\bf(A)$, and ${{\bf{A}}^H}$ denote its trace, rank, transpose, conjugate, and conjugate transpose respectively, while ${\bf{A}} \succeq {\bf{0}}$ means that $\bf{A}$ is a positive semidefinite matrix.
The operator $\textrm{vec}( \cdot )$ stacks the elements of a matrix in one long column vector, ${\textrm{invp}}( x )$ denotes the inverse of the positive portion, i.e., $\frac{1}{{\max ( {x,0} )}}$. $\|  \cdot  \|$, $(\cdot)!$, and $|  \cdot  |$ denote the Euclidean norm of a complex vector, the factorial operator, and the absolute value of a complex scalar, respectively.
${\mathbb{C}^{m \times n}}\;({\mathbb{R}^{m \times n}})$ denotes the space of ${m \times n}$ complex (real) matrices, and $\mathbb{R}_+\;(\mathbb{R}_-)$ denotes the set of positive (negative) real numbers. Finally, the symbol $\otimes$ denotes the Kronecker product of two vectors/matrices.

\section{System Model and Problem Formulation} \label{syste-mmodel}
\subsection{Proposed System Model}
\label{Section2:system}

\begin{figure}[hbtp]
  \centering
  \includegraphics[width = 0.48\textwidth]{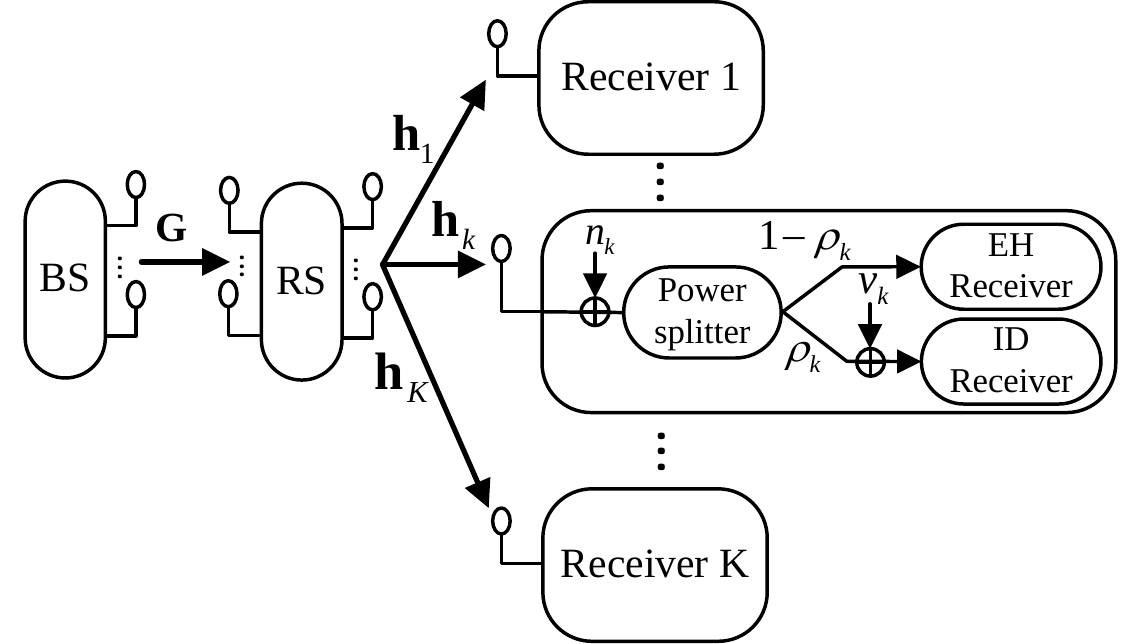}
  \caption{The multiuser MISO relay system with PS-based receiver. Each receiver splits the received signal into two parts for ID and EH, respectively.}\label{systemmodel}
\end{figure}
In this work, we consider a multiuser MISO relay system which consists of one BS, one RS, and $K$ mobile receivers indexed by $k \in \mathcal{K} \triangleq \{1,\ldots,K\}$. The number of antennas at the BS and the RS is denoted as $N_t$ and $N_r$, respectively, while each receiver is equipped with a single antenna. We assume that $K \leq \min{\{N_t,N_r\}}$, which provides sufficient degrees of freedom for signal detection. We also assume that the classical two-hop AF relaying protocol \cite{Laneman2001protocol} is employed, and that the direct links between the BS and the receivers are sufficiently weak (to be ignored). Different from the conventional multiuser MISO relay channels \cite{zhang2009joint}, we here consider PS-based receivers. The received signal at each receiver is split into two separate signal streams with different power levels: one is sent to the EH receiver and the other one is diverted to the ID receiver \cite{MIMOBroadcastingfor}, as shown in Fig. \ref{systemmodel}. Under the above assumptions, the signals are transmitted in two phases as explained below.

In the first phase, the BS transmits $K$ data streams, each carrying an independent message intended for one of the $K$ receivers. Thus, the transmitted data vector at the BS can be expressed as
\begin{equation}
{{\bf{x}}_B} = \sum\limits_{k = 1}^K {{{\bf{f}}_k}{s_k}} ,
\end{equation}
where $s_k$ is the data signal for receiver $k$, with zero mean and variance $E\{ {{{| {{s_k}} |}^2}} \} = 1$, and ${{\bf{f}}_k} \in {\mathbb{C}^{{N_t} \times 1}}$ denotes the transmit beamforming vector. The data signals $\{ s_k \}_{k \in \mathcal{K}}$ are assumed to be independent of each other. The transmit power of the BS can be shown as
\begin{equation}
{P_B} = E\{ {{{\bf{x}}_B}{\bf{x}}_B^H} \} = \sum\limits_{k = 1}^K {{{\| {{{\bf{f}}_k}} \|}^2}}.
\end{equation}
The transmission from the BS to the RS can be modeled as a standard point-to-point MIMO channel. Hence, the received data vector at the RS can be expressed as
\begin{equation}
{{\bf{y}}_R} = {\bf{G}}\sum\limits_{k = 1}^K {{{\bf{f}}_k}{s_k}}  + {{\bf{n}}_r},
\end{equation}
where ${\bf{G}} \in {\mathbb{C}^{{N_r} \times {N_t}}}$ denotes the MIMO channel from the BS to the RS, and ${{\bf{n}}_r} \in {\mathbb{C}^{{N_r} \times 1}}$ is the complex circular Gaussian noise vector at the RS, with zero mean and covariance $E\{ {{{\bf{n}}_r}{\bf{n}}_r^H} \} = \sigma _r^2{\bf{I}}$, where $\sigma _r^2$ is the average noise power. It is assumed that the transmitted signals $\{ s_k \}_{k \in \mathcal{K}}$ are independent of the noise vector ${{\bf{n}}_r}$.

In the second phase, the RS forwards the received signal to all the receivers after performing linear AF processing. Hence the vector signal transmitted from the RS can be formulated as
\begin{equation}
{{\bf{x}}_R} = {\bf{W}}{{\bf{y}}_R},
\end{equation}
where ${\bf{W}} \in {\mathbb{C}^{{N_r} \times {N_r}}}$ is the AF transformation matrix at the RS. The transmission power of the RS can be shown as
\begin{equation}
{P_R} = E\{ {{{\bf{x}}_R}{\bf{x}}_R^H} \} = \sum\limits_{k = 1}^K {{{\| {{\bf{WG}}{{\bf{f}}_k}} \|}^2}}  + \sigma _r^2{\| {\bf{W}} \|^2}.
\end{equation}
Finally, The signal received at the $k$th receiver, $k \in \mathcal{K}$, is given by
\begin{equation}
{y_k} = {\bf{h}}_k^H{\bf{WG}}\sum\limits_{k = 1}^K {{{\bf{f}}_k}{s_k}}  + {\bf{h}}_k^H{\bf{W}}{{\bf{n}}_r} + {n_k},
\end{equation}
where ${\bf{h}}_k^{} \in {\mathbb{C}^{{N_r} \times 1}}$ denotes the complex conjugate channel vector between the RS and receiver $k$, and ${n_k}$ is the additive noise introduced by the receive antenna at receiver $k$, which is assumed to be a complex circular Gaussian variable with zero mean and variance $\sigma _k^2$.

Let ${\rho _k}\left( {0 \le {\rho _k} \le 1} \right)$ denotes the PS ratio for receiver $k$, which means that portion ${\rho _k}$ of the signal power is used for signal detection while the remaining portion $1-{\rho _k}$ is diverted to an energy harvester. Thus, on the one hand, the signal available for ID at receiver $k$ can be expressed as
\begin{equation}
{\bf{r}}_k^{\textrm{ID}} = \sqrt {{\rho _k}} \left( {{\bf{h}}_k^H{\bf{WG}}\sum\limits_{k = 1}^K {{{\bf{f}}_k}{s_k}}  + {\bf{h}}_k^H{\bf{W}}{{\bf{n}}_r} + {n_k}} \right) + {v_k},
\end{equation}
where $v_k$ is the additional complex circular Gaussian circuit noise with zero mean and variance $E\{|v_k|^2\}=\omega_k^2$, resulting from phase offsets and non-linearities during baseband conversion \cite{MIMOBroadcastingfor}. Thus, the SINR at the $k$th receiver can be expressed as
\begin{equation} \label{nonrobust-SINR}
{\Gamma _k} = \frac{{{\rho _k}|{\bf{h}}_k^H{\bf{WG}}{{\bf{f}}_k}{|^2}}}{{{\rho _k}\left( {\sum\limits_{j \ne k}^K {|{\bf{h}}_k^H{\bf{WG}}{{\bf{f}}_j}{|^2}}  + \sigma _r^2\|{\bf{h}}_k^H{\bf{W}}\|{^2} + \sigma _k^2} \right) + \omega _k^2}}.
\end{equation}

On the other hand, the total harvested energy that can be stored by receiver $k$ is equal to
\begin{equation} \label{nonrobust-EH}
P_k^{\textrm{EH}} = {\xi _k}(1 - {\rho _k})\left( {\sum\limits_{j = 1}^K {{{| {{\bf{h}}_k^H{\bf{WG}}{{\bf{f}}_j}} |}^2}}  + \sigma _r^2{{\| {{\bf{h}}_k^H{\bf{W}}} \|}^2} + \sigma _k^2} \right),
\end{equation}
where ${\xi _k} \in \left( {0,1} \right]$ denotes the energy conversion efficiency of the $k$th EH unit, which indicates that only a portion ${\xi _k}$ of the radio frequency energy can be stored.

\subsection{Channel Error Model} \label{channel_model}


We assume that the required CSI can be estimated at the BS and the RS by means of a suitable channel estimation algorithms \cite{Ting2011Optimal,Sun2011Channel,Fan-Shuo2013Robust}. In this work, we consider a time division duplex (TDD) mode and assume that the radio channels vary sufficiently slowly over time, so that the downlink transmit CSI can be obtained by channel reciprocity \cite{Vook2004Signaling}. Therefore, the BS can obtain the MIMO channel matrix $\bf{G}$ involved the first phase of transmission, while the RS can obtain the channel vectors $\{{\bf{h}}_k\}$ involved in the second. However, the second phase CSI needs to be fed back from the RS to the BS using signaling channels \cite{Vook2004Signaling, Chan2008Coordinated, Yunlong2011Switched} for implementation of the design algorithms, as explained earlier.
Hence, here we assume that the BS can perfectly know the first phase CSI, matrix $\bf{G}$, but not the second phase CSI, i.e. vectors $\{{\bf{h}}_k\}$.
We consider quantization errors during channel feedback from the RS to the BS and employ the NBE model \cite{Relayprecoder} to characterize this type of imperfection. In particular, the true (but unknown) second phase CSI can be expressed as follows:
\begin{equation} \label{true-channel}
{{\bf{h}}_{k}} = {\widehat {\bf{h}}_{k}} + {{\bf{e}}_{k}},\;k \in \mathcal{K},
\end{equation}
where ${\widehat {\bf{h}}_{k}}$ denotes the estimated channel vector, while ${{\bf{e}}_{k}}$ denotes the CSI error vector. We assume that vector ${{\bf{e}}_{k}}$ is bounded in its Euclidean norm, that is
\begin{equation} \label{channelerrorvector}
\| {{{\bf{e}}_{k}}} \| \le {\eta _{k}},{\mkern 1mu} {\mkern 1mu} \;k \in \mathcal{K},
\end{equation}
where ${\eta _{k}}$ is a known positive constant. Equivalently, ${{\bf{h}}_{k}}$ belongs to the uncertainty set ${\Re _{k}}$ defined as
\begin{equation} \label{channelerror}
{\Re _{k}} = \{ {{\bf{h}}| {{\bf{h}} = {{\widehat {\bf{h}}}_{k}} + {{\bf{e}}_{k}}},\| {{{\bf{e}}_{k}}} \| \le {\eta _{k}}} \},\;k \in \mathcal{K}.
\end{equation}

The shape and the size of $\Re _{k}$ model the type of uncertainty in the estimated CSI, which is linked to the physical phenomenon producing the CSI errors. It should be emphasized that the actual errors ${{\bf{e}}_{k}}$ are assumed to be unknown while the corresponding upper bounds ${\eta _{k}}$ can be obtained using preliminary knowledge about the type of imperfection and/or coarse knowledge of the channel type and main characteristics \cite{Robustdownlinkpower}.

\subsection{Problem Formulation} \label{Problem-formulation}
In this subsection, we formulate the optimization problem for the joint design of $\bf{W}$, $\{{\bf{f}}_k\}_{k \in \mathcal{K}}$ and $\{\rho _k\}_{k \in \mathcal{K}}$ so as to minimize the total power consumption at the BS and the RS under the constraint that a set of minimum SINR and EH targets be satisfied at the receivers. In this study, we consider both non-robust and robust designs (against quantization errors).\footnote{The non-robust joint design is only based on the estimated CSI, while the robust design algorithm takes the CSI error into account.} The non-robust optimization problem can be expressed as\footnote{There is in general a power consumption tradeoff between the BS and the RS \cite{zhang2009joint}, and the objective function should be ${P_B} + \alpha {P_R}$ where $\alpha $ is a positive weight. We assume that $\alpha =1$ in this work, since this does not change the nature of the problem.}
\begin{equation} \label{nonrobust_problem}
\begin{array}{l}
\mathop {\min }\limits_{ {\bf{W}},\;\{{{\bf{f}}_k},\;{\rho _k}\} } \;{P_B} + {P_R}\\
{\textrm{s.t.}}\;{\Gamma _k} \ge {\gamma _k},\;P_k^{\textrm{EH}} \ge {\psi _k},\\
0 \le {\rho _k} \le 1,\;\forall k \in \mathcal{K},
\end{array}
\end{equation}
where in the evaluation of (\ref{nonrobust-SINR}) and (\ref{nonrobust-EH}), the true channel vectors ${\bf{h}}_k$ are replaced by their estimates $\widehat{{\bf{h}}}_k$.

Similarly, the robust optimization problem can be formulated as
\begin{equation} \label{robust_problem}
\begin{array}{l}
\mathop {\min }\limits_{{\bf{W}},\;\{{{\bf{f}}_k},\;{\rho _k}\} } \;{P_B} + {P_R}\\
{\textrm{s}}.{\textrm{t}}.\;{\Gamma _k} \ge {\gamma _k},\;P_k^{\textrm{EH}} \ge {\psi _k},\\
0 \le {\rho _k} \le 1,\; \|{{\bf{e}}_k}\|^2 \le \eta _k^2,\;\forall k  \in \mathcal{K},
\end{array}
\end{equation}
where in this case, the channel vectors in (\ref{nonrobust-SINR}) and (\ref{nonrobust-EH}) are given by (\ref{true-channel}).

It is not difficult to see that both (\ref{nonrobust_problem}) and (\ref{robust_problem}) are in general non-convex because both their objective functions and constraints are not convex over $\bf{W}$, $\{{\bf{f}}_k\}$ and $\{\rho _k\}$. Furthermore, (\ref{robust_problem}) involves an infinite number of constraints. These characteristics make it intractable to obtain the global optimal solution for (\ref{nonrobust_problem}) and (\ref{robust_problem}). In the sequel, we present two algorithms for obtaining suboptimal solutions to the above problems by applying proper convex optimization techniques, which are based on AO and SR, respectively. We complete this section with the following lemma.

\newtheorem{lem}{\underline{Lemma}}
\begin{lem} \label{lamma1}
Problem (\ref{nonrobust_problem}) is feasible for any finite user SINR targets if $\textrm{rank}({\bf{HG}}) = K$, where ${\bf{H}} = {[{{\bf{h}}_1}, \ldots ,{{\bf{h}}_K}]^H}$.
\end{lem}

\emph{Proof:} The feasibility of (\ref{nonrobust_problem}) is not connected to the EH constraints and PS ratios according to Lemma 3.1 \& Lemma 3.2 in \cite{JointTransmitBeamforming}. Hence, according to \emph{Theorem 1} in \cite{zhang2009joint}, we can easily see that the above lemma holds.

On the other hand, the feasibility of problem (\ref{robust_problem}) has not been well studied in the literature and still remains an open question, which would be an interesting topic for future research.

\section{Alternating Optimization Based Joint Transceiver Design} \label{AO}
In this section, we present the AO-based transceiver design algorithms for the jointly optimization of the BS beamforming vectors, the RS AF transformation matrix and the receiver PS ratios, for both non-robust and robust cases. In the proposed design, $\{ {{\bf{f}}_k},{\rho _k}\}_{k \in \mathcal{K}} $ and ${\bf{W}}$ are successively optimized in turn with the other fixed. We show that each subproblem for the optimization of $\{ {{\bf{f}}_k},{\rho _k}\} $ or ${\bf{W}}$ can be reformulated as an SDP problem based on the celebrated SDR technique and the S-procedure. Furthermore, modified randomization techniques based on the worst-case concept are provided to recover a rank-one solution when higher-rank solutions are returned.
\subsection{Non-Robust AO-Based Joint Transceiver Design} \label{nonrobust-AO}
In the following, we introduce a non-robust joint transceiver  design algorithm. First, let us consider the optimization of $\{ {{\bf{f}}_k},{\rho _k}\} $ in (\ref{nonrobust_problem}) while the RS AF matrix ${\bf{W}}$ is fixed. The celebrated SDR technique can be applied to solve the remaining optimization problem by introducing a new variable ${{\bf{F}}_k} = {{\bf{f}}_k}{\bf{f}}_k^H$. Hence, problem (\ref{nonrobust_problem}) can be reformulated as the following problem by ignoring the rank-one constraints for all ${{\bf{F}}_k}^\prime $s:
\begin{equation} \label{nonrobust-Fk}
\begin{array}{l}
\mathop {\min }\limits_{\{ {{\bf{F}}_k},\;{\rho _k}\} } \;\sum\limits_{k = 1}^K {\textrm{Tr}({{\bf{F}}_k})}  + \sum\limits_{k = 1}^K {\textrm{Tr}({{\bf{Q}}_k})}  + \sigma _r^2{\| {\bf{W}} \|^2}\\
\textrm{s.t.}\;\frac{1}{{{\gamma _k}}}{\widehat {\bf{h}}}_k^H{{\bf{Q}}_k}{{\widehat {\bf{h}}}_k} - \sum\limits_{j \ne k}^K {{\widehat {\bf{h}}}_k^H{{\bf{Q}}_j}{{\widehat {\bf{h}}}_k}}  \ge \sigma _r^2{\| {{\widehat {\bf{h}}}_k^H{\bf{W}}} \|^2} + \sigma _k^2 + \frac{{\omega _k^2}}{{{\rho _k}}},\\
\sum\limits_{j = 1}^K {{\widehat {\bf{h}}}_k^H{{\bf{Q}}_j}{{\widehat {\bf{h}}}_k}}  \ge \frac{{{\psi _k}}}{{{\xi _k}(1 - {\rho _k})}} - \sigma _k^2 - \sigma _r^2{\| {{\widehat {\bf{h}}}_k^H{\bf{W}}} \|^2},\\
{{\bf{F}}_k}\succeq{\bf{0}},\; 0 \le {\rho _k} \le 1,\;\forall k  \in \mathcal{K},
\end{array}
\end{equation}
where ${{\bf{Q}}_k} = {\bf{WG}}{{\bf{F}}_k}{{\bf{G}}^H}{{\bf{W}}^H}$. In this way, problem (\ref{nonrobust_problem}) is relaxed to a convex SDP problem, which can be efficiently solved by off-the-shelf algorithms \cite{Solvingsemidefinite}. Let $\{ {\bf{F}}_k^*\} $ and $\{ \rho _k^*\} $ denote the optimal solution to (\ref{nonrobust-Fk}). Based on \emph{Proposition 4.1} in \cite{JointTransmitBeamforming}, it can be verified that $\{ {\bf{F}}_k^*\} $ and $\{ \rho _k^*\} $ satisfy the first two sets of constraints of problem (\ref{nonrobust-Fk}) with equality and $\{ {\bf{F}}_k^*\} $ satisfy $\textrm{rank}({\bf{F}}_k^*) = 1,\;\forall k$. Thus the optimal solution of problem (\ref{nonrobust_problem}) can be expressed as $\{ {\bf{f}}_k^*,\rho _k^*\} $, where ${\bf{f}}_k^*$ is the principal component of ${\bf{F}}_k^*$, such that ${\bf{F}}_k^* = {\bf{f}}_k^*{\bf{f}}_k^{*H},\;\| {{\bf{f}}_k^*} \| = \sqrt {{f_k}} $ and $f_k$ is the largest eigenvalue of ${{\bf{F}}_k^*}$.

Next, we consider the optimization of the RS AF matrix ${\bf{W}}$ while assuming that $\{ {{\bf{f}}_k},{\rho _k}\} $ are fixed. Noting that ${{\bf{x}}^T\bf{Yz}} = \textrm{vec}{({{\bf{x}}}{{\bf{z}}^T})^T}\textrm{vec}({\bf{Y}})$, the numerator in (\ref{nonrobust-SINR}) can be rewritten as
\begin{equation} \label{nonrobust-nomi}
\begin{array}{l}
{\rho _k}{| {{\widehat {\bf{h}}}_k^H{\bf{WG}}{{\bf{f}}_k}} |^2} = {\rho _k}{| {{\bf{g}}_{kk}^T{\textrm{vec}}({\bf{W}})} |^2}\\
 = {\rho _k}{\bf{g}}_{kk}^T\widetilde {\bf{W}}\textrm{conj}({{\bf{g}}_{kk}}),
\end{array}
\end{equation}
where ${{\bf{g}}_{kj}} = {\textrm{vec}}({\textrm{conj}}({{\widehat {\bf{h}}}_k}){\bf{f}}_j^T{{\bf{G}}^T})$ and  $\widetilde {\bf{W}} = \textrm{vec}({\bf{W}})\textrm{vec}{({\bf{W}})^H}$. Similarly, the denominator in (\ref{nonrobust-SINR}) can be reformulated as
\begin{equation} \label{nonrobust-denomi}
\begin{array}{l}
{\rho _k}\sum\limits_{j \ne k}^K {{\bf{g}}_{kj}^T\widetilde {\bf{W}}\textrm{conj}({{\bf{g}}_{kj}})} \\
 + {\rho _k}\sigma _r^2\sum\limits_{j = 1}^{N_r} {{{{\widehat {\bf{h}}}_k^H}}{{\bf{E}}_j}\widetilde {\bf{W}}{\bf{E}}_j^H{{\widehat{\bf{h}}}_k}}  + {\rho _k}\sigma _k^2 + \omega _k^2,
\end{array}
\end{equation}
where ${{\bf{E}}_k} \in {\{ {0,1} \}^{{N_r} \times N_r^2}}$ is a linear mapping matrix such that ${{{\widehat{\bf{h}}}_k^H}}{{\bf{E}}_j}\textrm{vec}({\bf{W}}) = {\widehat{\bf{h}}}_k^H{\bf{W}}(:,j)$, where ${\bf{W}}(:,j)$ denotes the $j$th column of ${\bf{W}}$. With the help of (\ref{nonrobust-nomi}) and (\ref{nonrobust-denomi}), the SINR and EH constraints in problem (\ref{nonrobust_problem}) can be expressed as the following two inequalities
\begin{equation} \label{nonrobust_SINR}
\begin{array}{l}
\frac{1}{{{\gamma _k}}}{\bf{g}}_{kk}^T\widetilde {\bf{W}}\textrm{conj}({{\bf{g}}_{kk}}) - \sum\limits_{j \ne k}^K {{\bf{g}}_{kj}^T\widetilde {\bf{W}}\textrm{conj}({{\bf{g}}_{kj}}) - } \\
\sigma _r^2\sum\limits_{j = 1}^{N_r} {{{{{\widehat{\bf{h}}}_k^H}}}{{\bf{E}}_j}\widetilde {\bf{W}}{\bf{E}}_j^H{{\widehat{\bf{h}}}_k}}  \ge \sigma _k^2 + \frac{{\omega _k^2}}{{{\rho _k}}},
\end{array}
\end{equation}
\begin{equation} \label{nonrobust_EH}
\begin{array}{l}
\sum\limits_{j = 1}^K {{\bf{g}}_{kj}^T\widetilde {\bf{W}}\textrm{conj}({{\bf{g}}_{kj}})}  + \\
\sigma _r^2\sum\limits_{j = 1}^{N_r} {{{{{\widehat{\bf{h}}}_k^H}}}{{\bf{E}}_j}\widetilde {\bf{W}}{\bf{E}}_j^H{{\widehat{\bf{h}}}_k^H}}  \ge \frac{{{\psi _k}}}{{{\xi _k}(1 - {\rho _k})}} - \sigma _k^2.
\end{array}
\end{equation}
Hence, (\ref{nonrobust_problem}) can be reformulated as the following SDP problem by employing the well-known SDR technique
\begin{equation} \label{nonrobust-W}
\begin{array}{l}
\mathop {\min }\limits_{\widetilde {\bf{W}}} \;\sum\limits_{k = 1}^K \left({{\| {{{\bf{f}}_k}} \|}^2} + \textrm{Tr}({{\bf{C}}_k}\widetilde {\bf{W}}{\bf{C}}_k^H)\right)  + \sigma _r^2\textrm{Tr}(\widetilde {\bf{W}})\\
\textrm{s.t.}\; (\ref{nonrobust_SINR})\; \textrm{and}\; (\ref{nonrobust_EH}),\; \widetilde {\bf{W}}\succeq {\bf{0}},\;\forall k  \in \mathcal{K},
\end{array}
\end{equation}
where ${{\bf{C}}_k} = {({\bf{G}}{{\bf{f}}_k})^T} \otimes {{\bf{I}}_{{N_r}}}$. Different from (\ref{nonrobust-Fk}), the optimal solution ${\widetilde {\bf{W}}^*}$ to problem (\ref{nonrobust-W}) is not necessarily rank-one; hence a simple rank-one recovery method based on a randomization procedure is proposed in Appendix \ref{rankonerecovery} to address this issue.

The AO-based iterative algorithm to solve problem (\ref{nonrobust_problem}) is summarized in Table \ref{Non-robustAO}. Note that in each iteration of this algorithm, the objective function can only be decreased\footnote{Due to the rank-one recovery method, the total transmission power of step 2.2 is not necessarily less than that of step 2.1. However, we can terminate the algorithm if that happens to ensure the convergence of the algorithm. Note that this rare situation does not happen in all our simulations.} and it is also lower bounded by zero, thus the convergence of the iterative algorithm is guaranteed.

\begin{table}
  \centering
  \caption{AO-based non-robust transceiver design algorithm} \label{Non-robustAO}
  \begin{tabular}{p{0.9\columnwidth}}
 \hline
 \begin{itemize}
 \item[1.] Initialize ${\bf{W}}$ and define the tolerance of accuracy $\delta$.
 \item[2.] \textbf{Repeat}
  \begin{itemize}
 \item[2.1]  Solve problem (\ref{nonrobust-Fk}) with fixed ${\bf{W}}$ to obtain the updated $\{ {{\bf{f}}_k},{\rho _k}\} $.
 \item[2.2]  Solve problem (\ref{nonrobust-W}) with fixed $\{ {{\bf{f}}_k},{\rho _k}\} $ to obtain the updated ${\bf{W}}$. Employ the proposed rank-one recovery method in Appendix \ref{rankonerecovery} if higher-rank solutions are returned by solving problem (\ref{nonrobust-W}).
 \end{itemize}
 \item[3.] {\textbf{Until}} the total power consumption between two adjacent iterations is less than $\delta$ or the total power consumption of step $2.2$ is higher than that of step $2.1$.
 \end{itemize} \\
  \hline
 \end{tabular}
\end{table}

\subsection{Robust AO-Based Joint Transceiver Design} \label{robust-AO}
In this subsection, we address the robust counterpart of the non-robust AO-based joint transceiver design by employing the channel error model considered in Subsection \ref{channel_model}. We demonstrate that the corresponding subproblems can be converted to alternative forms where the concepts of SDR and S-procedure can be applied.

First, let us consider the optimization of $\{ {\bf{f}}_k,\rho_k\}$ in (\ref{robust_problem}) when $\bf{W}$ is fixed under imperfect CSI , i.e. ${{\bf{h}}_k} = {\widehat {\bf{h}}_k} + {{\bf{e}}_k},\;\forall k \in \mathcal{K}$. In this case, the SINR constraints in (\ref{nonrobust-Fk}) can be replaced by the following quadratic forms:
\begin{equation} \label{robust_SINR}
{({\widehat {\bf{h}}_k} + {{\bf{e}}_k})^H}{{\bf{U}}_k}({\widehat {\bf{h}}_k} + {{\bf{e}}_k}) \ge \sigma _k^2 + \frac{{\omega _k^2}}{{{\rho _k}}},
\end{equation}
where ${{\bf{e}}_k}$ satisfy (\ref{channelerrorvector}) and ${{\bf{U}}_k}$ can be expressed as
\begin{equation}
{{\bf{U}}_k} = \frac{1}{{{\gamma _k}}}{{\bf{Q}}_k} - \sum\limits_{j \ne k}^K {{{\bf{Q}}_j}}  - \sigma _r^2{\bf{W}}{{\bf{W}}^H}.
\end{equation}
Similarly, the EH constraints can be reformulated as the following expression:
\begin{equation} \label{robust_EH}
{({\widehat {\bf{h}}_k} + {{\bf{e}}_k})^H}{{\bf{V}}_k}({\widehat {\bf{h}}_k} + {{\bf{e}}_k}) + \sigma _k^2 \ge \frac{{{\psi _k}}}{{{\xi _k}(1 - {\rho _k})}},\forall \|{{\bf{e}}_k}\|^2 \le \eta _k^2,
\end{equation}
where ${{\bf{V}}_k} = \sum\limits_{j = 1}^K {{{\bf{Q}}_j}}  + \sigma _r^2{\bf{W}}{{\bf{W}}^H}$.

By applying the S-procedure, the constraints in (\ref{robust_SINR}) and (\ref{robust_EH}) can be reformulated as finite convex constraints, which are equivalent to the following two linear matrix inequality (LMI) constraints
\begin{equation} \label{robust_SINR1}
\begin{array}{l}
\left[ {\begin{array}{*{20}{c}}
{{{\bf{U}}_k} + {\lambda _k}{\bf{I}}}&{{{\bf{U}}_k}{{\widehat {\bf{h}}}_k}}\\
{\widehat {\bf{h}}_k^H{{\bf{U}}_k}}&{\widehat {\bf{h}}_k^H{{\bf{U}}_k}{{\widehat {\bf{h}}}_k} - \sigma _k^2 - \omega _k^2{p _k} - {\lambda _k}\eta _k^2}
\end{array}} \right]\succeq{\bf{0}},
\end{array}
\end{equation}
\begin{equation} \label{robust_EH1}
\begin{array}{*{20}{l}}
{\left[ {\begin{array}{*{20}{c}}
{{{\bf{V}}_k} + {\mu _k}{\bf{I}}}&{{{\bf{V}}_k}{{\widehat {\bf{h}}}_k}}\\
{\widehat {\bf{h}}_k^H{{\bf{V}}_k}}&{\widehat {\bf{h}}_k^H{{\bf{V}}_k}{{\widehat {\bf{h}}}_k} + \sigma _k^2 - \frac{{{\psi _k}}}{{{\xi _k}}}{q _k} - {\mu _k}\eta _k^2}
\end{array}} \right]\succeq{\bf{0}},}
\end{array}
\end{equation}
where ${p _k} = \frac{1}{{\rho {}_k}}$, ${q _k} = \frac{1}{{1 - \rho {}_k}}$, while ${\lambda _k} \geq 0$ and $\mu_k \geq 0$ are slack variables.

With (\ref{robust_SINR1}) and ($\ref{robust_EH1}$), the robust counterpart of problem (\ref{nonrobust-Fk}) can be reformulated as follows by ignoring the rank-one constraints for all ${{\bf{F}}_k}^\prime $s:
\begin{equation} \label{robust-Fk}
\begin{array}{*{20}{l}}
{\mathop {\min }\limits_{\{ {{\bf{F}}_k},\;{p_k},\;{q_k},\;{\lambda _k},\;{\mu _k}\}} \;\sum\limits_{k = 1}^K {{\textrm{Tr}}({{\bf{F}}_k})}  + \sum\limits_{k = 1}^K {{\textrm{Tr}}({{\bf{Q}}_k})}  + \sigma _r^2{{\| {\bf{W}} \|}^2}}\\
{{\textrm{s}}.{\textrm{t}}.\;} (\ref{robust_SINR1}) \; \textrm{and} \; (\ref{robust_EH1}),
{\lambda _k} \ge 0,\;{\mu _k} \ge 0,\\
{p_k} \ge 1,\;{q_k} \ge 1,\;\textrm{invp} (p_k) + \textrm{invp} (q_k) \le 1,\\
{{{\bf{F}}_k}\succeq{\bf{0}},\;\forall k  \in \mathcal{K}}.
\end{array}
\end{equation}
It is worth noting that different from problem (\ref{nonrobust-Fk}) which always returns a rank-one solution, the optimal solution $\{ {\bf{F}}_k^*\} $ of (\ref{robust-Fk}) may not be of rank-one, in which case, additional processing steps may be needed to extract a rank-one solution from the ${\bf{F}}_k^*$. To this end, a rank-one recovery method based on a randomization procedure but with lower complexity is presented in Appendix \ref{rankonerecovery}.

Next, we consider the optimization of the RS AF matrix $\bf{W}$ with fixed $\{{\bf{f}}_k,\rho_k\}$. Similar to Subsection \ref{nonrobust-AO}, we have the following expression
\begin{align}
{{\bf{g}}_{kj}} &= {\textrm{vec}}({\textrm{conj}}({{\bf{h}}_k}){\bf{f}}_j^T{{\bf{G}}^T}) = ({\bf{G}}{{\bf{f}}_j} \otimes {{\bf{I}}_{{N_r}}}){\textrm{conj}}({{\bf{h}}_k})\\
 &= {\widehat {\bf{G}}_j}{\textrm{conj}}({{\bf{h}}_k}),\notag
\end{align}
where ${\widehat {\bf{G}}_j} = {\bf{G}}{{\bf{f}}_j} \otimes {{\bf{I}}_{{N_r}}}$, $j \in \mathcal{K}$. Thus, the robust version of (\ref{nonrobust_SINR}) can be formulated as
\begin{equation} \label{robust_WSINR}
{\begin{array}{*{20}{l}}
{\frac{1}{{{\gamma _k}}}(\widehat {\bf{h}}_k^H + {\bf{e}}_k^H){{\overline {\bf{W}} }_k}({{\widehat {\bf{h}}}_k} + {{\bf{e}}_k})}\\
{ - \sum\limits_{j \ne k}^K {(\widehat {\bf{h}}_k^H + {\bf{e}}_k^H){{\overline {\bf{W}} }_j}({{\widehat {\bf{h}}}_k} + {{\bf{e}}_k})} }\\
{ - \sigma _r^2\sum\limits_{j = 1}^{N_r} {(\widehat {\bf{h}}_k^H + {\bf{e}}_k^H){{\bf{E}}_j}\widetilde {\bf{W}}{\bf{E}}_j^H({{\widehat {\bf{h}}}_k} + {{\bf{e}}_k})} }\\
{ \ge \sigma _k^2 + \frac{{\omega _k^2}}{{{\rho _k}}},\;\forall \|{{\bf{e}}_k}\|^2 \le \eta _k^2,\; k \in \mathcal{K}},
\end{array}}
\end{equation}
where ${\overline {\bf{W}} _j} = \widehat {\bf{G}}_j^T\widetilde {\bf{W}}{\textrm{conj}}({\widehat {\bf{G}}_j})$, and $\widetilde {\bf{W}}$ has already been defined in Subsection \ref{nonrobust-AO}.
By applying the S-procedure, we can transform (\ref{robust_WSINR}) into the following LMIs
\begin{equation} \label{robust_SINR2}
\begin{array}{*{20}{l}}
{\left[ {\begin{array}{*{20}{c}}
{{{\widetilde {\bf{U}}}_k} + {\lambda _k}{\bf{I}}}&{{{\widetilde {\bf{U}}}_k}{{\widehat {\bf{h}}}_k}}\\
{\widehat {\bf{h}}_k^H{{\widetilde {\bf{U}}}_k}}&{\widehat {\bf{h}}_k^H{{\widetilde {\bf{U}}}_k}{{\widehat {\bf{h}}}_k} - \sigma _k^2 - \omega _k^2{p_k} - {\lambda _k}\eta _k^2}
\end{array}} \right]\succeq{\bf{0}},}
\end{array}
\end{equation}
where ${\widetilde {\bf{U}}_k} = \frac{1}{{{\gamma _k}}}{\overline {\bf{W}} _k} - \sum\limits_{j \ne k}^K {{{\overline {\bf{W}} }_j}}  - \sigma _r^2\sum\limits_{k = 1}^{{N_r}} {{{\bf{E}}_k}\widetilde {\bf{W}}{\bf{E}}_k^H} $ and ${\lambda _k}\geq 0$.
Similarly, the robust version of (\ref{nonrobust_EH}) can be formulated as
\begin{equation} \label{robust_EH2}
\begin{array}{l}
\left[ {\begin{array}{*{20}{c}}
{{{\widetilde {\bf{V}}}_k} + {\mu _k}{\bf{I}}}&{{{\widetilde {\bf{V}}}_k}{{\widehat {\bf{h}}}_k}}\\
{\widehat {\bf{h}}_k^H{{\widetilde {\bf{V}}}_k}}&{\widehat {\bf{h}}_k^H{{\widetilde {\bf{V}}}_k}{{\widehat {\bf{h}}}_k} + \sigma _k^2 - \frac{{{\psi _k}}}{{{\xi _k}}}{q_k} - {\mu _k}\eta _k^2}
\end{array}} \right]\succeq{\bf{0}},
\end{array}
\end{equation}
where ${\widetilde {\bf{V}}_k} = \sum\limits_{j = 1}^K {{{\overline {\bf{W}} }_k}}  + \sigma _r^2\sum\limits_{k = 1}^{{N_r}} {{{\bf{E}}_k}\widetilde {\bf{W}}{\bf{E}}_k^H} $ and ${\mu _k}\geq 0$.
Hence, problem (\ref{robust_problem}) with fixed $\{ {\bf{f}}_k,\rho_k\}$ can be formulated as the following SDP problem with the SDR technique
\begin{equation} \label{robust-W}
\begin{array}{*{20}{l}}
{\mathop {\min }\limits_{\widetilde {\bf{W}},\;\{ {\lambda _k},\;{\mu _k}\} } \;\sum\limits_{k = 1}^K \left({{{\left\| {{{\bf{f}}_k}} \right\|}^2} + {\textrm{Tr}}({{\bf{C}}_k}\widetilde {\bf{W}}{\bf{C}}_k^H)}\right)  + \sigma _r^2{\textrm{Tr}}(\widetilde {\bf{W}})}\\
{{\textrm{s}}.{\textrm{t}}.\;\;(\ref{robust_SINR2})\; \textrm{and}\; (\ref{robust_EH2}),\;\widetilde {\bf{W}}\succeq {\bf{0}},\;{\lambda _k} \ge 0,\;{\mu _k} \ge 0,\;\forall k},
\end{array}
\end{equation}
where ${{\bf{C}}_k}$ has been defined in Subsection \ref{nonrobust-AO}. Note that the optimal solution of (\ref{robust-W}) is also not necessarily rank-one and thus we can employ the method in Appendix \ref{rankonerecovery} to address this issue.

We summarize the AO-based robust joint transceiver design algorithm in Table \ref{RobustAO}. For practical implementation, the optimally designed AF matrix $\bf{W}$ and PS ratios $\{\rho_k\}$ should be fed forward from the BS to the RS and receivers, respectively, through signaling channels prior to data transmission.

\begin{table}
  \centering
  \caption{AO-based robust transceiver design algorithm} \label{RobustAO}
  \begin{tabular}{p{0.9\columnwidth}}
 \hline
 \begin{itemize}
 \item[1.] Initialize ${\bf{W}}$ and define the tolerance of accuracy $\delta$.
 \item[2.] \textbf{Repeat}
  \begin{itemize}
 \item[2.1]  Solve problem (\ref{robust-Fk}) with fixed ${\bf{W}}$ to obtain the updated $\{ {{\bf{F}}_k},{\rho _k}\}$. If $\{{{\bf{F}}_k}\}$ are rank-one matrices, use their principal eigenvectors as a solution. Otherwise, employ the rank-one recovery method to recover a rank-one solution from $\{{{\bf{F}}_k}\}$.
 \item[2.2]  Solve problem (\ref{robust-W}) with fixed $\{ {{\bf{F}}_k},{\rho _k}\}$ to obtain the updated ${\bf{W}}$. If $ \widetilde {\bf{W}}$ is a rank-one matrix, use its principal eigenvector as a solution. Otherwise, employ the method in Appendix \ref{rankonerecovery} to recover a rank-one solution from $ \widetilde {\bf{W}}$.
 \end{itemize}
 \item[3.] {\textbf{Until}} the total power consumption between two adjacent iterations is less than $\delta$ or increased.
 \end{itemize} \\
  \hline
 \end{tabular}
\end{table}

\section{Switched Relaying Based Joint Transceiver Design} \label{SR}
\begin{figure*}[hbtp]
  \centering
  \includegraphics[width = 0.7\textwidth]{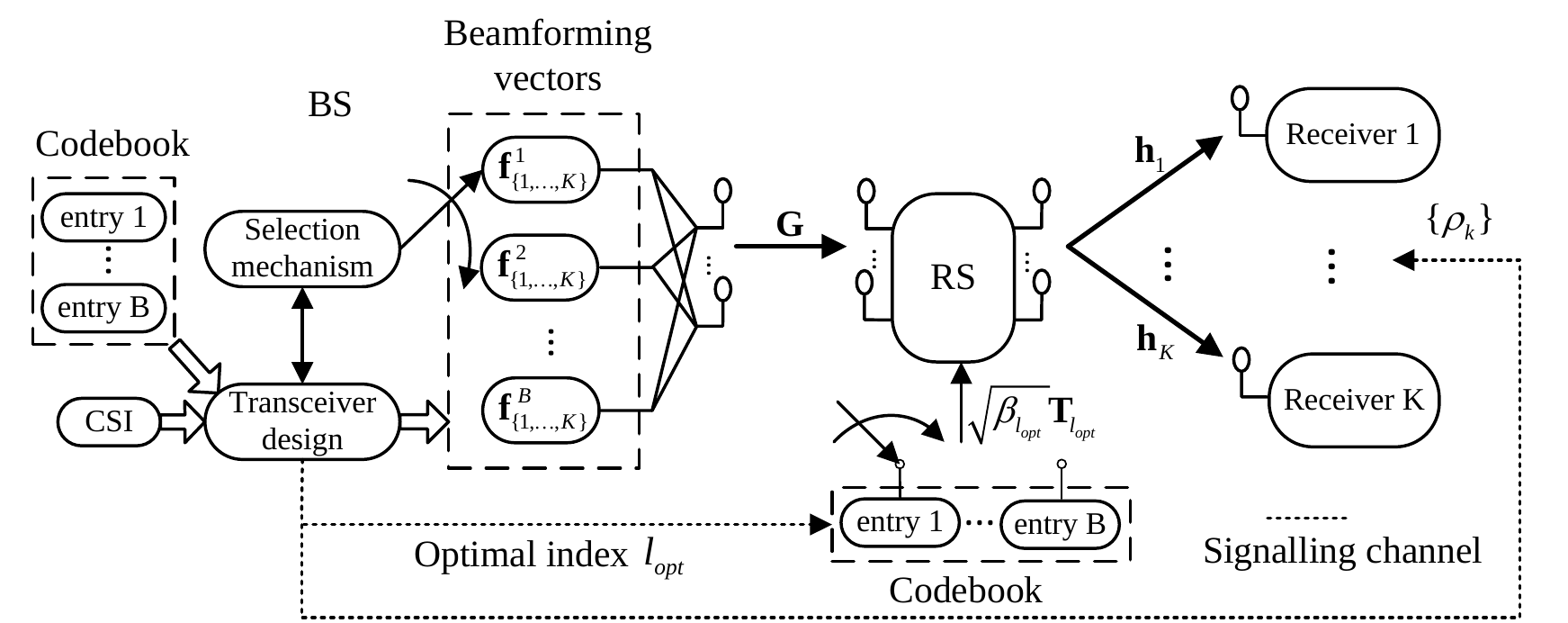}
  \caption{Proposed transceiver scheme with switched relaying processing for multiuser MISO relay system. The dashed line denotes signaling channels.}\label{switched_relaying}
\end{figure*}

In the previous section, we proposed novel AO-based transceiver design algorithms. As will be shown in Section \ref{Simulation}, these algorithms can achieve an excellent performance in transmission, but they are characterized by higher signaling overhead and computational complexity. In this section, motivated by these considerations, we develop alternative SR-based transceiver design algorithms that are more efficient and simpler to implement. As illustrated in Fig. \ref{switched_relaying}, we equip the RS with a finite codebook of permutation matrices\footnote{A permutation matrix is a square binary matrix that has exactly one entry equal to $1$ in each row and each column, while all the other entries are equal to $0$.}, i.e., $\Upsilon  = \{ {{\bf{T}}_1},{{\bf{T}}_2}, \ldots ,{{\bf{{ T}}}_{B}}\} $, where ${\bf{T}}_l \in \{0,1\}^{N_r \times N_r}$ is the $l$-th matrix in the codebook, index $l \in \{1,\ldots,B\}$ and ${B}$ denotes the codebook size which satisfies ${B}\ll {N_r}!$.\footnote{The total number of permutation matrices at the RS is ${N_r}!$. It is not realistic to use all the permutation matrices as the codebook when $N_r$ is large.
Thus, we propose to construct a codebook of permutation matrices with  $B$ elements, ${B}\ll {N_r}!$. The codebook design approach will be introduced in Subsection \ref{codebookdesign}. In practice, the value of $B$ should be chosen to achieve a suitable tradeoff between performance requirements and implementation complexity.} In order to reduce the signaling overhead and the number of optimization variables, the RS AF matrix $\bf{W}$ is constructed by multiplying the appropriate permutation matrix from the codebook with a power scaling factor.
That is to say, the optimization of ${\bf{W}}$ is replaced by ${\sqrt{\beta_l}} {{\bf{T}}_l}$, where ${\beta_l}$ is a variable power scaling factor. Thus, each permutation matrix gives rise to a permuted channel matrix for which we can design a so-called \emph{latent} transceiver, consisting of the BS beamforming vectors, the RS power scaling factor and the receiver PS ratios. Among the $B$ latent transceivers so designed, by an algorithm to be developed below, the optimal one with index $l_{opt}$ is chosen by means of a suitable selection criterion for transmission. Specifically, the selection mechanism is designed to choose the optimal latent transceiver with the minimum power consumption, which can be expressed as
\begin{equation} \label{selection}
{l_{opt}} = \arg \mathop {\min \,}\limits_l {P_l},
\end{equation}
where ${P_l}$ denotes the total transmission power corresponding to the $l$th latent transceiver.

Before data transmission, the BS only sends the index of the permutation matrix, the RS power scaling factor and the receiver PS ratios corresponding to the optimum transceiver to the RS and the receivers. Compared to the AO-based algorithms, the SR-based algorithms can significantly reduce the number of signaling bits and the computational complexity. The proposed SR-based scheme works as follows.
\begin{itemize}
\item The BS designs the $B$ latent transceivers based on available permutation matrices within the codebook; it then determines the optimal latent transceiver based on (\ref{selection}).
\item The BS sends the optimal permutation index $l_{opt}$, the corresponding RS power scaling factor, and the PS ratios corresponding to the optimal transceiver to the RS and the receivers through signaling channels.
\item The RS constructs the optimal RS AF matrix based on the forwarded RS power scaling factor and the stored permutation matrix with index $l_{opt}$.
\end{itemize}

We introduce the non-robust and robust latent transceiver design algorithms in Subsection \ref{nonrobustSR} and \ref{robustSR}, respectively. In Subsection \ref{simplified_SR}, a simplified SR-based transceiver design algorithm is proposed. The design approach for the codebook of permutation matrices is presented in Subsection \ref{codebookdesign}. 

\subsection{Non-Robust Latent Transceiver Design} \label{nonrobustSR}

Firstly, we introduce a non-robust algorithm to construct the $l$th latent transceiver. We aim to design the beamforming vectors $\{{\bf{f}}_k^l\}$, the RS power scaling factor $\beta_l$ and the receiver PS ratios $\{{\rho_k^l}\}$ so as to minimize the sum of BS and RS transmit power under both SINR and EH constraints. We note that superscripts $l$ in $\{{\bf{f}}_k^l\}$, $\{{\rho_k^l}\}$, etc., have been removed in the following for notational simplicity. For each latent transceiver, the optimization problem can be formulated as
\begin{equation} \label{nonrobust_switched}
\begin{array}{*{20}{l}}
\mathop {\min }\limits_{\{ {{{\bf{f}}_k},\;{\rho _k}} \},\;{\beta _l}} \;{P_B} + {P_R}\\
{{\textrm{s}}.{\textrm{t}}.\;{\Gamma _k} \ge {\gamma _k},\;P_k^{\textrm{EH}} \ge {\psi _k},}\\
{0 \le {\rho _k} \le 1,\;{\bf{W}} = \sqrt {{\beta _l}} {{\bf{T}}_l},\;\forall k  \in \mathcal{K}.}
\end{array}
\end{equation}
As we can see, although we replaced the AF matrix $\bf{W}$ with $\sqrt{{\beta_l}}{\bf{T}}_l$, problem (\ref{nonrobust_switched}) is still non-convex and difficult to solve due to the coupling between variables ${{\bf{f}}_k}$, ${\rho _k}$ and ${\beta _l}$. Our proposed method is motivated by the observation that problem (\ref{nonrobust_switched}) can be reformulated as a difference of convex (DC) programming problem with proper transformations. Thus, the concept of CCCP \cite{yuille2003concave, lanckriet2009convergence} can be adopted to iteratively solve the DC problem, as explained below.

First, we introduce the variable substitution
\begin{equation}
{\varphi _l} = \frac{1}{{{\beta _l}}},
\end{equation}
and further define the following vectors:
\begin{equation} \label{definition_vector}
\begin{array}{l}
{\bf{p}} = {[{p_1}, \ldots ,{p_K}]^T},\;{\bf{q}} = {[{q_1}, \ldots ,{q_K}]^T},\\
{\bf{f}} = {[{\bf{f}}_1^T, \ldots ,{\bf{f}}_K^T]^T},{\bf{r}} = [{\bf{p}}^T,{\bf{q}}^T,{\varphi _l},{\bf{f}}^T]^T,
\end{array}
\end{equation}
where $p_k$ and $q_k$ have already been introduced in Subsection \ref{robust-AO}, alongside with (\ref{robust_SINR1}) and (\ref{robust_EH1}).
Then, the objective function of problem (\ref{nonrobust_switched}) can be transformed into
\begin{equation}
P_l({\bf{r}}) = \sum\limits_{k = 1}^K {{\bf{f}}_k^H{{\bf{f}}_k}}  + \sum\limits_{k = 1}^K {\frac{{{\bf{f}}_k^H{{\bf{G}}^H}{\bf{G}}{{\bf{f}}_k}}}{{{\varphi _l}}}}  + \sigma _r^2\frac{1}{{{\varphi _l}}},
\end{equation}
which is strictly jointly convex in the variables $\{ {\varphi _l},{{\bf{f}}}\}  \in {\mathbb{R}_ + } \times \mathbb{C}{}^{K{N_t} \times 1}$ \cite{ConvexOptimization}.

Similar to the transformation of the objective function, the SINR constraints can be rewritten as
\begin{equation}
{w_k}( {\bf{r}})  - {x_k}( {\bf{r}})  \le 0,
\end{equation}
where ${w_k}( {\bf{r}})$ and ${x_k}( {\bf{r}})$ are defined as
\begin{align} \label{CCCP_1}
{w_k}( {\bf{r}}) &= \sum\limits_{j \ne k}^K {{\bf{f}}_j^H{{\bf{G}}^H}{\bf{T}}_l^H{{\widehat{\bf{h}}}_k}{\widehat{\bf{h}}}_k^H{{\bf{T}}_l}{\bf{G}}{{\bf{f}}_j}} \\
&+ \sigma _k^2{\varphi _l} + \sigma _r^2{\widehat{\bf{h}}}_k^H{{\widehat{\bf{h}}}_k} + \frac{1}{4}\omega _k^2{({p_k} + {\varphi _l})^2}, \notag
\end{align}
\begin{equation} \label{CCCP_2}
{x_k}( {\bf{r}})  = \frac{1}{4}\omega _k^2{({p_k} - {\varphi _l})^2} + \frac{1}{{{\gamma _k}}}{\bf{f}}_k^H{{\bf{G}}^H}{\bf{T}}_l^H{{\widehat{\bf{h}}}_k}{\widehat{\bf{h}}}_k^H{{\bf{T}}_l}{\bf{G}}{{\bf{f}}_k}.
\end{equation}
Moreover, the EH constraints can be recast as
\begin{equation}
{y_k}( {\bf{r}})  - {z_k}( {\bf{r}})  \le 0,
\end{equation}
where
\begin{equation} \label{CCCP_3}
{y_k}( {\bf{r}})  = \frac{{{\psi _k}}}{{4{\xi _k}}}{({q_k} + {\varphi _l})^2} - \sigma _r^2{\widehat{\bf{h}}}_k^H{{\widehat{\bf{h}}}_k},
\end{equation}
\begin{align} \label{CCCP_4}
{z_k}( {\bf{r}}) & = \sum\limits_{j = 1}^K {{\bf{f}}_j^H{{\bf{G}}^H}{\bf{T}}_l^H{{\widehat{\bf{h}}}_k}{\widehat{\bf{h}}}_k^H{{\bf{T}}_l}{\bf{G}}{{\bf{f}}_j}} \\
 &+ \sigma _k^2{\varphi _l} + \frac{{{\psi _k}}}{{4{\xi _k}}}{({q_k} - {\varphi _l})^2}. \notag
\end{align}
We remark that (\ref{CCCP_1}), (\ref{CCCP_2}), (\ref{CCCP_3}) and (\ref{CCCP_4}) are all convex functions jointly with respect to the variables in ${\bf{r}} \in \mathbb{R}_+^K \times \mathbb{R}_+^K \times \mathbb{R}_+ \times \mathbb{C}^{KN_t \times 1}$. Thus, problem (\ref{nonrobust_switched}) can be equivalently reformulated as the following DC program:
\begin{equation} \label{nonrobust-CCCP}
\begin{array}{l}
{\mathop {\min }\limits_{\bf{r}} \;P_l({\bf{r}})}\\
{\textrm{s}}.{\textrm{t}}.\;{w_k}({\bf{r}}) - {x_k}({\bf{r}}) \le 0,\;{y_k}({\bf{r}}) - {z_k}({\bf{r}}) \le 0,\\
{p_k} \ge 1,\;{q_k} \ge 1,\;{\textrm{invp}}({p_k}) + {\textrm{invp}}({q_k}) \le 1,\;\forall k  \in \mathcal{K},
\end{array}
\end{equation}
where the last set of inequality constraints must be satisfied with equality at optimality, for otherwise, the objective value can be further decreased by decreasing the ${p_k}$'s.

According to the concept of CCCP, we approximate the functions ${x_k}({\bf{r}})$ and ${z_k}({\bf{r}})$ in the $i$th iteration by their first-order Taylor expansions around the current point ${{\bf{r}}^{(i)}}$, denoted as ${\widehat x_k}({{\bf{r}}^{(i)}},{\bf{r}})$ and ${\widehat z_k}({{\bf{r}}^{(i)}},{\bf{r}})$, respectively. With the help of \cite{brandwood1983complex} and \cite{cheng2012joint}, ${\widehat x_k}({{\bf{r}}^{(i)}},{\bf{r}})$ is given by
\begin{equation} \label{affineapprox}
{\widehat x_k}({{\bf{r}}^{(i)}},{\bf{r}}) = {x_k}({{\bf{r}}^{(i)}}) + 2\mathcal{R}\{ \nabla {x_k}{({{\bf{r}}^{(i)}})^H}({\bf{r}} - {{\bf{r}}^{(i)}})\},
\end{equation}
where $\nabla {x_k}{({{\bf{r}}^{(i)}})}$ denotes the conjugate derivative of the function ${x_k}({{\bf{r}}})$ with respect to the complex vector ${{\bf{r}}}$.\footnote{Since ${{\bf{r}}}$ is composed of both real and complex variables, we make a modification of $\nabla {x_k}{({{\bf{r}}^{(i)}})}$ such that (\ref{affineapprox}) holds for ${{\bf{r}}}$.} We note that ${\widehat x_k}({{\bf{r}}^{(i)}},{\bf{r}})$ is an affine function of ${{\bf{r}}}$. $\nabla {x_k}{({{\bf{r}}^{(i)}})}$ is given by
\begin{equation}
\begin{array}{l}
\nabla {x_k}({{\bf{r}}^{(i)}}) = [{{\bf{0}}_{1 \times (k - 1)}},{\mkern 1mu} {\kern 1pt} \frac{1}{4}\omega _k^2(p_k^{(i)} - \varphi _l^{(i)}),{\mkern 1mu} {\kern 1pt} {{\bf{0}}_{1 \times (2K - k)}},{\mkern 1mu} {\kern 1pt} \\
 - \frac{1}{4}\omega _k^2(p_k^{(i)} - \varphi _l^{(i)}),{\mkern 1mu} {\kern 1pt} \frac{1}{{{\gamma _k}}}{({{\bf{G}}^H}{\bf{T}}_l^H{\widehat {\bf{h}}_k}\widehat {\bf{h}}_k^H{{\bf{T}}_l}{\bf{Gf}}_k^{(i)})^T}{]^T}.
\end{array}
\end{equation}
Similarly, ${\widehat z_k}({{\bf{r}}^{(i)}},{\bf{r}})$ can be expressed as
\begin{equation}  \label{affineapproz}
{\widehat z_k}({{\bf{r}}^{(i)}},{\bf{r}}) = {z_k}({{\bf{r}}^{(i)}}) + 2\mathcal{R}\{ \nabla {z_k}{({{\bf{r}}^{(i)}})^H}({\bf{r}} - {{\bf{r}}^{(i)}})\},
\end{equation}
where
\begin{equation}
\begin{array}{l}
\nabla {z_k}({{\bf{r}}^{(i)}}) = [{\bf{0}}_{1\times(K+k-1)},\;\frac{{{\psi _k}}}{{4{\xi _k}}}(q_k^{(i)} - \varphi _l^{(i)}),\;{\bf{0}}_{1\times(K-k)},\\
\frac{1}{2}\sigma _k^2 - \frac{{{\psi _k}}}{{4{\xi _k}}}(q_k^{(i)} - \varphi _l^{(i)}),\;
{({\widetilde {\bf{G}}_k}{\bf{f}}_1^{(i)})^T}, \ldots ,{({\widetilde {\bf{G}}_k}{\bf{f}}_K^{(i)})^T}{]^T},\\
{\widetilde {\bf{G}}_k} = {{\bf{G}}^H}{\bf{T}}_l^H{{\widehat{\bf{h}}}_k}{\widehat{\bf{h}}}_k^H{{\bf{T}}_l}{\bf{G}}.
\end{array}
\end{equation}

Then, in the $i$th iteration of the proposed CCCP based algorithm, we have the following convex optimization problem
\begin{equation} \label{nonrobust-CCCP1}
\begin{array}{*{20}{l}}
{\mathop {\min }\limits_{\bf{r}} \;P_l({\bf{r}})}\\
{{\textrm{s}}.{\textrm{t}}.\;{w_k}({\bf{r}}) - {{\widehat x}_k}({{\bf{r}}^{(i)}},{\bf{r}}) \le 0,\;{y_k}({\bf{r}}) - {{\widehat z}_k}({{\bf{r}}^{(i)}},{\bf{r}}) \le 0,}\\
{{p_k} \ge 1,\;{q_k} \ge 1,\;{\textrm{invp}}({p_k}) + {\textrm{invp}}({q_k}) \le 1,}\;\forall k  \in \mathcal{K},
\end{array}
\end{equation}
whose solution is denoted by ${{\bf{r}}^{(i + 1)}}$.

\newtheorem{lemma}{\underline{Proposition}}
\begin{lemma}
By introducing a new set of variables $d_k$, $\widetilde d_k$, $e_k$ and $\widetilde e_k$, $k \in \mathcal{K}$, problem (\ref{nonrobust-CCCP1}) can be reformulated as the following SOCP problem
\begin{equation} \label{nonrobust-CCCP-SOCP}
\begin{array}{*{20}{l}}
{\mathop {\min }\limits_{{\bf{r}},\;{P_1},\;{P_2},\;{P_3}} \;{P_1} + {P_2} + {P_3}}\\
{\textrm{s}}.{\textrm{t}}.\;\|[{\bf{f}}_1^T, \ldots ,{\bf{f}}_K^T,({P_1} - 1)/2]\| \le ({P_1} + 1)/2,\\
\| [{({\bf{G}}{{\bf{f}}_1})^T}, \ldots ,{({\bf{G}}{{\bf{f}}_K})^T},({P_2} - {\varphi _l})/2] \| \le ({P_2} + {\varphi _l})/2,\\
\left\| [2{\sigma _r},{\varphi _l} - {P_3}]\right\| \le {\varphi _l} + {P_3},\\\vspace{+0.1em}
\left\| {\left[ {{\bf{w}}_k^T,\frac{{\widetilde d_k - {d_k} - 1}}{2}} \right]} \right\| \le \frac{{\widetilde d_k - {d_k} + 1}}{2},\forall k,\\
\left\| {\left[ {\sqrt {\frac{{{\psi _k}}}{{4{\xi _k}}}} ({q_k} + {\varphi _l}),\frac{{\widetilde e_k - {e_k} - 1}}{2}} \right]} \right\| \le \frac{{\widetilde e_k - {e_k} + 1}}{2},\\
{{p_k} \ge 1,\;{q_k} \ge 1,\;{\textrm{invp}}({p_k}) + {\textrm{invp}}({q_k}) \le 1,\forall k \in \mathcal{K}},
\end{array}
\end{equation}
where ${\bf{w}}_k$ is defined in (\ref{definitionw}).
\end{lemma}

\emph{Proof:} Please refer to Appendix \ref{appendixB}.

We can show that the proposed CCCP based iterative algorithm for the $l$th transceiver design converges to a local optimal solution of problem (\ref{nonrobust_switched}). The proof is similar to that of \emph{Lemma 2} and \emph{Theorem 1} in \cite{cheng2012joint}, and we therefore omit the details.
We summarize the proposed non-robust latent transceiver design algorithm in Table \ref{Non-robustSR}.
\begin{table}
  \centering
  \caption{Non-robust latent transceiver design algorithm} \label{Non-robustSR}
  \begin{tabular}{p{0.9\columnwidth}}
 \hline
 \begin{itemize}
 \item[1.] Define the tolerance of accuracy $\delta$ and the maximum number of iteration $N_{max}$.
 \item[2.] \textbf{For} $l = 1, \ldots ,B$
 \item[3.] \begin{itemize} \item Initialize the algorithm with a feasible point ${\bf{r}}^{(0)}$ which is obtained by solving problem (\ref{nonrobust_switched}) with ${\bf{W}} = {\bf{T}}_l$. Set the iteration number $i=0$.  \end{itemize}
 \item[4.] \begin{itemize}
 \item \textbf{Repeat}
 \item  Compute the affine approximation ${\widehat x_k}({{\bf{r}}^{(i)}},{\bf{r}})$ and ${\widehat z_k}({{\bf{r}}^{(i)}},{\bf{r}})$ according to (\ref{affineapprox}) and (\ref{affineapproz}), respectively.
 \item  Solve problem (\ref{nonrobust-CCCP-SOCP}), and assign the solution to ${\bf{r}}^{(i+1)}$.
 \item  Update the iteration number : $i=i+1$.
 \item \textbf{Until} $| P_l({\bf{r}}^{(i+1)}) - P_l({\bf{r}}^{(i)}) | \le \delta $ or the maximum number of iterations is reached, i.e., $i > {N_{\max }}$.
 \end{itemize}
 \item[5.] \begin{itemize} \item Obtain the $l$th latent transceiver $\{{\bf{f}}_k^*, \rho_k^*,\beta_l^*\}$. \end{itemize}
 \item[6.] \textbf{End} 
 \end{itemize} \\
  \hline
 \end{tabular}
\end{table}

\subsection{Robust Latent Transceiver Design } \label{robustSR}
Secondly, let us propose a robust latent transceiver design algorithm by taking the channel errors into consideration. As we can see, the concept of SR can also be applied to problem (\ref{robust_problem}), which can be reformulated as follows:
\begin{equation} \label{robust_switched}
\begin{array}{*{20}{l}}
{\mathop {\min }\limits_{\{ {{{\bf{f}}_k},\;{\rho _k}} \},\;{\beta _l}} \;{P_B} + {P_R}}\\
{{\textrm{s}}.{\textrm{t}}.\;{\Gamma _k} \ge {\gamma _k},\;P_k^{\textrm{EH}} \ge {\psi _k},\;0 \le {\rho _k} \le 1},\\
{{\bf{W}} = \sqrt {{\beta _l}} {{\bf{T}}_l},\; \|{{\bf{e}}_k}\|^2 \le \eta _k^2,\;\forall k  \in \mathcal{K}.}
\end{array}
\end{equation}
We employ a subgradient-type iterative algorithm to solve this problem. Our proposed method is motivated by the observation that, if we fix ${\beta _l}$, then problem (\ref{robust_switched}) is equivalent to problem (\ref{robust-Fk}) with the SDR technique.\footnote{The concept of CCCP can also be applied to the robust problem (\ref{robust_switched}). However, based on simulations we find that the performance of the CCCP based algorithm is inferior to that of the subgradient-type algorithm. Thus, we employ the subgradient-type algorithm instead.} With fixed $\beta_l$, (\ref{robust_SINR1}) and (\ref{robust_EH1}) can be reformulated as the following two LMIs:
\begin{equation} \label{robust_SINR3}
\begin{array}{*{20}{l}}
{\left[ {\begin{array}{*{20}{c}}
{{{\overline {\bf{U}} }_k} + {\lambda _k}{\bf{I}}}&{{{\overline {\bf{U}} }_k}{{\widehat {\bf{h}}}_k}}\\
{\widehat {\bf{h}}_k^H{{\overline {\bf{U}} }_k}}&{\widehat {\bf{h}}_k^H{{\overline {\bf{U}} }_k}{{\widehat {\bf{h}}}_k} - \frac{{\sigma _k^2}}{{{\beta _l}}} - \frac{{\omega _k^2{p _k}}}{{{\beta _l}}} - {\lambda _k}\eta _k^2}
\end{array}} \right]\succeq{\bf{0}},}
\end{array}
\end{equation}
\begin{equation} \label{robust_EH3}
\begin{array}{*{20}{l}}
{\left[ {\begin{array}{*{20}{c}}
{{{\overline {\bf{V}} }_k} + {\mu _k}{\bf{I}}}&{{{\overline {\bf{V}} }_k}{{\widehat {\bf{h}}}_k}}\\
{\widehat {\bf{h}}_k^H{{\overline {\bf{V}} }_k}}&{\widehat {\bf{h}}_k^H{{\overline {\bf{V}} }_k}{{\widehat {\bf{h}}}_k} + \frac{{\sigma _k^2}}{{{\beta _l}}} - \frac{{{\psi _k}}}{{{\xi _k}{\beta _l}}}{q _k} - {\mu _k}\eta _k^2}
\end{array}} \right]\succeq{\bf{0}},}
\end{array}
\end{equation}
where ${\overline {\bf{U}} _k} = \frac{1}{{{\gamma _k}}}{\widetilde {\bf{Q}}_k} - \sum\limits_{j \ne k}^K {{{\widetilde {\bf{Q}}}_j}}  - \sigma _r^2{\bf{I}}$, ${\overline {\bf{V}} _k} = \sum\limits_{j = 1}^K {{{\widetilde {\bf{Q}}}_j}}  + \sigma _r^2{\bf{I}}$, and ${\widetilde {\bf{Q}}_k} = {{\bf{T}}_l}{\bf{G}}{{\bf{F}}_k}{{\bf{G}}^H}{\bf{T}}_l^H$.

Then, problem (\ref{robust_switched}) with the use of the SDR technique and fixed $\beta_l$ can be expressed as
\begin{equation} \label{robust_fixedbeta}
\begin{array}{*{20}{l}}
\begin{array}{l}
f({\beta _l}) = \\
\mathop {\min }\limits_{\{ {{\bf{F}}_k},\;{p_k},\;{q_k},\;{\lambda _k},\;{\mu _k}\} } \;\sum\limits_{k = 1}^K {{\textrm{Tr}}({{\bf{F}}_k})}  + {\beta _l}\sum\limits_{k = 1}^K {{\textrm{Tr}}({{\widetilde {\bf{Q}}}_k})}  + \sigma _r^2{\beta _l}
\end{array}\\
{{\textrm{s}}.{\textrm{t}}.\;(\ref{robust_SINR3})\; \textrm{and} \;(\ref{robust_EH3}) ,\;{\lambda _k} \ge 0,\;{\mu _k} \ge 0,}\\
{p_k} \ge 1,\;{q_k} \ge 1,\;\textrm{invp} (p_k) + \textrm{invp} (q_k) \le 1,\\
{{{\bf{F}}_k}\succeq{\bf{0}},\;\forall k  \in \mathcal{K}.}\\
\end{array}
\end{equation}
We here consider the SDR version of problem (\ref{robust_switched}) instead of its original form with rank-one constraints, since strong duality holds if problem (\ref{robust_fixedbeta}) is feasible. The partial dual problem of (\ref{robust_fixedbeta}) can be formulated as (\ref{robust_switched_dual}) shown at the top of next page, where ${{\bf{X}}_k} = \left[ {\begin{array}{*{20}{c}}
{{{\overline {\bf{X}} }_k}}&{{{\bf{x}}_k}}\\
{{\bf{x}}_k^H}&{{x_k}}
\end{array}} \right]$ and ${{\bf{Y}}_k} = \left[ {\begin{array}{*{20}{c}}
{{{\overline {\bf{Y}} }_k}}&{{{\bf{y}}_k}}\\
{{\bf{y}}_k^H}&{{y_k}}
\end{array}} \right]$ denote the dual variables associated with constraints (\ref{robust_SINR3}) and (\ref{robust_EH3}), respectively. It is worth noting that ${{\bf{X}}_k}$ and ${{\bf{Y}}_k}$ can
be obtained as side information provided with any standard SDP solver, since dual variables are served as a certificate for optimality.
\begin{figure*}
\begin{equation} \label{robust_switched_dual}
\begin{array}{*{20}{l}}
{f({\beta _l}) = \mathop {\max }\limits_{\{ {{\bf{X}}_k},\;{{\bf{Y}}_k}\} } \;\mathop {\min }\limits_{\{ {{\bf{F}}_k},\;{\rho _k}\} } \; - \sum\limits_{k = 1}^K {Tr\left( {{{\bf{X}}_k}\left[ {\begin{array}{*{20}{c}}
{{{\overline {\bf{U}} }_k} + {\lambda _k}{\bf{I}}}&{{{\overline {\bf{U}} }_k}{{\widehat {\bf{h}}}_k}}\\
{\widehat {\bf{h}}_k^H{{\overline {\bf{U}} }_k}}&{\widehat {\bf{h}}_k^H{{\overline {\bf{U}} }_k}{{\widehat {\bf{h}}}_k} - \frac{{\sigma _k^2}}{{{\beta _l}}} - \frac{{\omega _k^2{\alpha _k}}}{{{\beta _l}}} - {\lambda _k}\eta _k^2}
\end{array}} \right]} \right)} }\\
{ - \sum\limits_{k = 1}^K {Tr\left( {{{\bf{Y}}_k}\left[ {\begin{array}{*{20}{c}}
{{{\overline {\bf{V}} }_k} + {\mu _k}{\bf{I}}}&{{{\overline {\bf{V}} }_k}{{\widehat {\bf{h}}}_k}}\\
{\widehat {\bf{h}}_k^H{{\overline {\bf{V}} }_k}}&{\widehat {\bf{h}}_k^H{{\overline {\bf{V}} }_k}{{\widehat {\bf{h}}}_k} + \frac{{\sigma _k^2}}{{{\beta _l}}} - \frac{{{\psi _k}}}{{{\xi _k}{\beta _l}}}{\beta _k} - {\mu _k}\eta _k^2}
\end{array}} \right]} \right)}  + \sum\limits_{k = 1}^K {{\textrm{Tr}}({{\bf{F}}_k})}  + {\beta _l}\sum\limits_{k = 1}^K {{\textrm{Tr}}({{\widetilde {\bf{Q}}}_k})}  + \sigma _r^2{\beta _l},}
\end{array}
\end{equation}
\hrulefill
\end{figure*}

Next, we can use the subgradient method \cite{Gan2012Generic} to iteratively solve the SDR version of problem (\ref{robust_switched}). At the $(i+1)$th iteration, the power scaling factor $\beta_l$ can be updated according to
\begin{equation} \label{robust_updatebeta}
{\beta _l}(i + 1) = [{\beta _l}(i) - \theta (i)s(i)]_\varepsilon ^ + ,
\end{equation}
where $[ \cdot ]_\varepsilon ^ +  = \max \left( {.,\varepsilon } \right)$, $\theta (i)$ is the step-size in the $i$th iteration and $s(i)$ denotes a subgradient of $f({\beta _l})$ at ${\beta _l}(i)$. The subgradient $s(i)$ can be calculated as \cite{Pennanen2014Decentralized}
\begin{align}
s(i) &= \sum\limits_{k = 1}^K {{\textrm{Tr}}(\widetilde {\bf{Q}}_k^*)}  + \sigma _r^2 - \sum\limits_{k = 1}^K {{x_k^*}( {\frac{{\sigma _k^2}}{{{\beta _l}{{(i)}^2}}} + \frac{{\omega _k^2}}{{\rho _k^*{\beta _l}{{(i)}^2}}}} )} \\
&+ \sum\limits_{k = 1}^K {{y_k^*}( {\frac{{\sigma _k^2}}{{{\beta _l}{{(i)}^2}}} - \frac{{{\psi _k}}}{{(1 - \rho _k^*){\xi _k}{\beta _l}{{(i)}^2}}}} )},\notag
\end{align}
where $\widetilde {\bf{Q}}_k^* = {{\bf{T}}_l}{\bf{GF}}_k^*{{\bf{G}}^H}{\bf{T}}_l^H$, $\{{\bf{F}}_k^*,\rho_k^*\}$ denotes the optimal solution of problem (\ref{robust_fixedbeta}) and $\{x_k^*,y_k^*\}$ denote the lower right corner elements of ${{\bf{X}}_k^*}$ and ${{\bf{Y}}_k^*}$, which are the optimal dual variables associated with (\ref{robust_SINR3}) and (\ref{robust_EH3}).

Finally, the robust latent transceiver design algorithm is summarized in Table \ref{robust-SR}.\footnote{For the non-robust case, we can also employ the subgradient method to address problem (\ref{nonrobust_switched}) with each subproblem being an SDP problem. It is well known that solving an SDP problem requires relatively high computational complexity compared with solving an SOCP problem. Moreover, it is important to note that the two algorithms for the non-robust case have very close performance in our simulations. Thus, we employ the CCCP based algorithm for the non-robust case.}
\begin{table}
  \centering
  \caption{Robust latent transceiver design algorithm} \label{robust-SR}
  \begin{tabular}{p{0.9\columnwidth}}
 \hline
 \begin{itemize}
 \item[1.] Initialize the step-size $\theta$, define the tolerance of accuracy $\delta$ and the maximum number of iterations $N_{max}$.
 \item[2.] \textbf{For} $l = 1, \ldots ,B$
 \item[3.] \begin{itemize} \item Initialize ${\beta _l}(0)$. Set the iteration number $i=0$. \end{itemize}
 \item[4.] \begin{itemize}
 \item \textbf{Repeat}
 \item  Solve problem (\ref{robust_fixedbeta}) to obtain the optimal $\{{\bf{F}}_k^*, \rho_k^*\}$ and the corresponding dual variables $\{x_k^*,y_k^*\}$.
 \item  Update the power scaling factor according to (\ref{robust_updatebeta}).
 \item  Update the iteration number : $i=i+1$.
 \item {\textbf{Until}} $\left| {f({\beta _l}(i + 1)) - f({\beta _l}(i))} \right| \le \delta $ or the maximum number of iterations is reached, i.e., $i > {N_{\max }}$.
 \end{itemize}
 \item[5.] \begin{itemize} \item Obtain the $l$th latent transceiver $\{{\bf{f}}_k^*, \rho_k^*,\beta_l^*\}$.\end{itemize}
 \item[6.] {\textbf{End}} 

 \end{itemize} \\
  \hline
 \end{tabular}
\end{table}
\subsection{Proposed Simplified SR-based Transceiver Design} \label{simplified_SR}
As shown in Subsection \ref{nonrobustSR} and \ref{robustSR}, the proposed SR-based transceiver design algorithm involves devising $B$ latent transceivers corresponding to the elements in $\Upsilon $ where for each transceiver, we employ iterative methods to address the highly non-convex problem. Specifically, the design of each transceiver involves an iterative algorithm to obtain the corresponding BS beamforming vectors, RS power scaling factor and receiver PS ratios. However, for given CSIs only the best transceiver with the minimum total power consumption is selected for transmission according to the selection mechanism (\ref{selection}) while the other $B-1$ solutions are discarded. Thus, there will be a considerable waste in the computational resources.
In order to improve the proposed SR-based transceiver design algorithm and make it more suitable for practical implementation, we propose a heuristic approach, referred to as the \emph{simplified} latent transceiver design algorithm, to address the aforementioned problem. Based on simulation experiments, we find that a good initial point obtained by solving problem (\ref{nonrobust_switched}) or (\ref{robust_switched}) with fixed initial power scaling factor $\beta_l$ almost always leads to a better convergent point than by solving with a not so good initial point. Thus, we only design the particular transceiver with the best initial point.
The proposed simplified non-robust latent transceiver design algorithm is summarized in Table \ref{SimplifiednonrobustSR}. The robust version of the simplified latent transceiver design algorithm is similar to the non-robust case, and thus is omitted here. We will show in Section \ref{Simulation} that the simplified SR-based transceiver design algorithms can achieve a similar performance with that of the proposed SR-based algorithms in Table \ref{Non-robustSR} and \ref{robust-SR}.
\begin{table}
  \centering
  \caption{Simplified non-robust latent transceiver design algorithm} \label{SimplifiednonrobustSR}
  \begin{tabular}{p{0.9\columnwidth}}
 \hline
 \begin{itemize}
 \item[1.] Define the tolerance of accuracy $\delta$ and the maximum iteration number $N_{max}$.
 \item[2.] \textbf{For} the $l$th latent transceiver ($l=1,\ldots,B$)
 \item[3.] \begin{itemize} \item Compute a initial feasible point ${\bf{r}}^{(0)}$ which is obtained by solving problem (\ref{nonrobust_switched}) with ${\bf{W}} = {\bf{T}}_l$.  \end{itemize}
 \item[4.] {\bf{end}}
 \item[5.] Choose the initial ${\bf{r}}^{(0)}$ and ${\bf{T}}_{{l_{opt}}}^{}$ with the minimum objective value. Set the iteration number $i=0$.
 \item[6.] \textbf{Repeat}
 \begin{itemize} \item Compute the affine approximation ${\widehat x_k}({{\bf{r}}^{(i)}},{\bf{r}})$ and ${\widehat z_k}({{\bf{r}}^{(i)}},{\bf{r}})$ according to (\ref{affineapprox}) and (\ref{affineapproz}), respectively.
  \item Solve problem (\ref{nonrobust-CCCP-SOCP}), and assign the solution to ${\bf{r}}^{(i+1)}$.
  \item Update the iteration number : $i=i+1$. \end{itemize}
 \item[7.] \textbf{Until} $| P_{l_{opt}}({\bf{r}}^{(i+1)}) - P_{l_{opt}}({\bf{r}}^{(i)}) | \le \delta $ or the maximum number of iterations is reached, i.e., $i > {N_{\max }}$.
 \end{itemize} \\
  \hline
 \end{tabular}
\end{table}


\subsection{Codebook Design} \label{codebookdesign}
In this subsection, we propose two algorithms to construct the codebook of permutation matrices.
The basic principle of the proposed algorithms is to choose the permutation matrices which are more likely to result in lower transmission power. We make a promising observation that when the singular values of the permuted channel matrix ${\bf{H}}{{\bf{T}}_l}{\bf{G}}$ are larger or more equally distributed, the total power consumption is usually smaller, where the channel matrix ${\bf{H}}$ has already been defined in Subsection \ref{Problem-formulation} and ${\bf{H}}{{\bf{T}}_l}{\bf{G}}$ represents the equivalent channel matrix between the BS and the receivers.

Let $ {\pi _k^{l}}$ denotes the $k$th singular value of ${\bf{H}}{{\bf{T}}_l}{\bf{G}}$, $k \in \mathcal{K}$. The first scheme (referred to as \emph{Sum-Max} method) constructs the codebook of permutation matrices by choosing the ones which correspond to the $B$ largest sums of singular values, i.e.,
\begin{equation} \label{select1}
\{ {{\bf{T}}_{{1}}}, \ldots ,{{\bf{T}}_{{B}}}\}  = \arg \mathop {{\mathop{\textrm {maxB}}\nolimits} }\limits_{{{\bf{T}}_l}} \;\left(\sum\limits_{k = 1}^K {\pi _k^l}\right) ,
\end{equation}
where ${{\mathop{\textrm {maxB}}\nolimits} }(\cdot)$ returns the permutation matrices correspond to the $B$ largest values of its argument.

Alternatively, the second scheme (referred to as \emph{Max-Min} method) chooses the permutation matrices to maximize the smallest singular value, i.e.,
\begin{equation} \label{select2}
\{ {{\bf{T}}_1}, \ldots ,{{\bf{T}}_B}\}  = \arg \mathop {{\textrm{maxB}}}\limits_{{{\bf{T}}_l}} \;\left( {\mathop {\min }\limits_{k \in \mathcal{K} } (\pi _k^l)} \right).
\end{equation}
It is worth noting that the proposed two schemes are heuristic techniques which may serve as a shortcut to the process of finding a satisfactory solution. The performance of the codebook design algorithms will be studied in the simulation results in Section \ref{Simulation}.

\section{Complexity Analysis} \label{sec-complexity}
In Section \ref{AO} and \ref{SR}, we proposed the AO-based and SR-based transceiver design algorithms for problem (\ref{nonrobust_problem}) and (\ref{robust_problem}), respectively. In this section, we compare the relative computational complexity of the proposed transceiver design algorithms. Moreover, we apply the same basic element of complexity analysis as in \cite{Wang2014Outage}.
Among the proposed algorithms in this paper, we consider:
\begin{itemize}
\item \emph{Non-robust AO:} the algorithm summarized in Table \ref{Non-robustAO}.
\item \emph{Robust AO:} the algorithm summarized in Table \ref{RobustAO}.
\item \emph{Non-robust SR:} the algorithm summarized in Table \ref{Non-robustSR}.
\item \emph{Robust SR:} the algorithm summarized in Table \ref{robust-SR}.
\item \emph{Simplified non-robust SR:} the algorithm summarized in Table \ref{SimplifiednonrobustSR}.
\item \emph{Simplified robust SR:} the robust counterpart of the algorithm in Table \ref{SimplifiednonrobustSR}.
\end{itemize}

The complexity of the \emph{non-robust AO} algorithm is dominated by solving problems (\ref{nonrobust-Fk}) and (\ref{nonrobust-W}) $I_1$ times , where $I_1$ denotes the iteration number. We note that the dual problems of (\ref{nonrobust-Fk}) and (\ref{nonrobust-W}) can be solved instead of (\ref{nonrobust-Fk}) and (\ref{nonrobust-W}) for better efficiency. Consider the dual problem of (\ref{nonrobust-Fk}), which involves $K$ LMI constraints of size $N_t$ and on the order of  $n_1=\mathcal{O}(KN_t^2 + K)$ decision variables. Thus, the complexity of a generic interior-point method for solving problem (\ref{nonrobust-Fk}) is given by $\mathcal{O}(n_1\sqrt {K{N_t}} ( {KN_t^3 + {n_1} KN_t^2 + {n_1^2}} ))$. Similarly, the complexity of solving the dual problem of (\ref{nonrobust-W}) can be written as $\mathcal{O}({n_2}\sqrt {N_r^2} ( {N_r^6 + {n_2}N_r^4 + n_2^2} ))$, where ${n_2} = \mathcal{O}(N_r^4)$ is the order of the corresponding decision variables. The complexity of solving problem (\ref{rank-one-recovery-2}) can be neglected since it admits closed-form solution. Thus, the overall complexity of the \emph{non-robust AO} algorithm is on the order of the quantity shown in the first row of Table \ref{complexity}.

The complexity of the \emph{robust AO} algorithm is dominated by solving problems (\ref{robust-Fk}) and (\ref{robust-W}) $I_2$ times, plus the complexity of the rank-one recovery method. Problem (\ref{robust-Fk}) involves $KN_t^2 + 4K$ variables, $2K$ LMI constraints of size $N_r + 1$ and $K$ LMI constraints of size $N_t$. Problem (\ref{robust-W}) involves $N_r^4+2K$ variables, $2K$ LMI constraints of size $N_r + 1$ and $1$ LMI constraint of size $N_r^2$. Moreover, the complexity of the rank-one recovery method in Table \ref{rank-one-recovery} is dominated by solving problem (\ref{rank-one-recovery-1}) $R$ times, where $R$ is the number of randomization steps.\footnote{The complexity to recover ${{{\bf{W}}^*}}$ is negligible since problem (\ref{rank-one-recovery-2}) has closed-form solution.} Problem (\ref{rank-one-recovery-1}) involves $2K$ variables and $2K$ second-order cone (SOC) constraints of dimension $3$. Thus, the overall complexity of the \emph{robust AO} algorithm is on the order of the quantity shown in the second row of Table \ref{complexity}.

The complexity of the \emph{non-robust SR} algorithm is dominated by solving problem (\ref{nonrobust-CCCP-SOCP}) $BI_3$ times, where $I_3$ is the iteration number, since the complexity of computing the affine approximation ${\widehat x_k}({{\bf{r}}^{(i)}},{\bf{r}})$ and ${\widehat z_k}({{\bf{r}}^{(i)}},{\bf{r}})$ is negligible compared to solving (\ref{nonrobust-CCCP-SOCP}). Problem (\ref{nonrobust-CCCP-SOCP}) involves $2K+3$ SOC constraints, including $1$ SOC of dimension $KN_t+2$, $1$ SOC of dimension $KN_r+2$, $K$ SOCs of dimension $K+2$, and $K+1$ SOCs of dimension $3$. The number of variables is on the order of $\mathcal{O}(KN_t+2K)$. It follows that the complexity of the \emph{non-robust SR} algorithm is on the order of the quantity shown in the third row of Table \ref{complexity}.

The complexity of the \emph{robust SR}, \emph{simplified non-robust SR} and \emph{simplified robust SR} algorithms can be analyzed in a similar way; the corresponding complexity figures are shown in the fourth to sixth rows of Table \ref{complexity}, respectively, where $I_4$ denotes the iteration number of the \emph{(simplified) robust SR} algorithm. It is of interest to investigate the asymptotic complexity of the proposed algorithms when $N_t$, $N_r$ and $K$ are large, i.e., when we let $N_r = N_t = K \rightarrow \infty $. We further assume that $I_1 = I_2 = I_3 = I_4 = I$ for simplicity. Under these conditions, one can verify that the complexities of the proposed algorithms in Table \ref{complexity} are on the orders of $2IN_t^{13}$, $2\sqrt{3}I N_t^{13}$, $6BIN_t^{6.5}$, $3\sqrt{3}BIN_t^{10}$, $6(B+I)N_t^{6.5}$ and $3\sqrt{3}(B+I)N_t^{10}$, respectively. As seen, the robust algorithms always consume more computational resources than their non-robust counterparts, the SR-based algorithms have lower complexity compared with the AO-based algorithms and the simplified SR-based algorithms have the lowest complexity.
\begin{table*}
  \centering
  \caption{Complexity analysis of the proposed transceiver designs} \label{complexity}
  \begin{tabular}{|c|c|c|c|c|c|c}
\hline \hline
   \multicolumn{1}{|c}{Proposed design} & \multicolumn{1}{|c|}{Complexity Order (suppressing the $\ln ( {{1 \mathord{\left/
 {\vphantom {1 \varepsilon }} \right.
 \kern-\nulldelimiterspace} \varepsilon }} )$ )}\\ \hline \hline
   \multirow{1}*{Non-robust AO} & {$I_1(\mathcal{O}(n_1\sqrt {K{N_t}} ( {KN_t^3 + {n_1} KN_t^2 + {n_1^2}} )) + \mathcal{O}({n_2}N_r ( {N_r^6 + {n_2}N_r^4 + n_2^2} )))$,}
    $n_1=\mathcal{O}(KN_t^2 + K)$, ${n_2} = \mathcal{O}(N_r^4)$\\ \hline
   \multirow{3}* {Robust AO} & {$I_2 ( {\cal O}( {{n_1}\sqrt {2K({N_r} + 1) + K{N_t}} ( {2K{{({N_r} + 1)}^3} + KN_t^3 + 2{n_1}K{{({N_r} + 1)}^2} + {n_1}KN_t^2 + n_1^2} )} )+$ }\\
   & ${\cal O}( {{n_2}\sqrt {2K({N_r} + 1) + N_r^2} ( {2K{{({N_r} + 1)}^3} + N_r^6 + 2{n_2}K{{({N_r} + 1)}^2} + {n_2}N_r^4 + n_2^2} )} )+$\\
   & $R({\cal O}( {2K\sqrt {4K}( {2K {3^2} + 4{K^2}} )} )))$, ${n_1} = \mathcal{O}(KN_t^2 + 4K)$, ${n_2} = \mathcal{O}(N_r^4 + 2K)$\\ \hline
   \multirow{1}*{Non-robust SR} & {$BI_3({\cal O}( {{n}\sqrt {4K+6}( (KN_t+2)^2 + (KN_r+2)^2 + K(K+2)^2+(K+1) 3^2 + n^2 )} ))$, $n=\mathcal{O}(KN_t+2K)$} \\
      \hline
   \multirow{2}*{Robust SR} & {$BI_4( {O( {{n_1}\sqrt {2K({N_r} + 1) + K{N_t}} ( {2K{{({N_r} + 1)}^3} + KN_t^3 + 2{n_1}K{{({N_r} + 1)}^2} + {n_1}KN_t^2 + n_1^2} )} )} )$} \\
   & $R({\cal O}( {2K\sqrt {4K}( {2K {3^2} + 4{K^2}} )} )))$, ${n_1} = \mathcal{O}(KN_t^2 + 4K)$. \\
   \hline
    \multirow{1}*{Simplified non-robust SR} & {$(B+I_3)({\cal O}( {{n}\sqrt {4K+6}( (KN_t+2)^2 + (KN_r+2)^2 + K(K+2)^2+(K+1) 3^2 + n^2 )} ))$, $n=\mathcal{O}(KN_t+2K)$} \\
      \hline
 \multirow{2}*{Simplified robust SR} & {$(B+I_4)( {O( {{n_1} \sqrt {2K({N_r} + 1) + K{N_t}} ( {2K{{({N_r} + 1)}^3} + KN_t^3 + 2{n_1}K{{({N_r} + 1)}^2} + {n_1}KN_t^2 + n_1^2} )} )} )$} \\
    & +$R({\cal O}( {2K \sqrt {4K}( {2K {3^2} + 4{K^2}} )} )))$, ${n_1} = \mathcal{O}(KN_t^2 + 4K)$. \\
   \hline
 \end{tabular}
\end{table*}

\section{Simulation Results} \label{Simulation}
In order to evaluate the performance of the proposed transceiver designs, numerical results have been obtained by performing computer simulations. In the simulations, we assume that both the first phase and the second phase channel coefficients are flat-fading i.i.d. with unit variance Rayleigh distribution. The nominal system configuration is defined by the following choice of parameters: ${N_t} = {N_r} = 4,\;K = 3$, $\xi = 1$, ${\sigma_k^2} = {\sigma ^2} = -30$dBm, ${\omega_k^2} = {\omega ^2} = -20$dBm, and ${\sigma_r ^2} = -30$dBm unless otherwise specified. In addition, we assume equal SINR and EH thresholds at the destination receivers, i.e., ${\gamma _k} = \gamma ,\;{\psi _k} = \psi ,\;\forall k \in \mathcal{K}$, and equal norm bounds for the second phase channel error vectors, i.e., ${\eta _{kj}} = \eta ,\;\forall j,k$ for simplicity. In the implementations of the various algorithms, the tolerance parameter is chosen as $\delta = 2 \times {10^{ - 3}}$ while $I_1=I_2=20$ and $I_3=I_4=50$; all the rank-one recovery methods employ $R = 100$ randomization steps. All convex problems are solved by CVX \cite{CVX} on a desktop Intel (i3-2100) CPU running at 3.1GHz with 4GB RAM.

\begin{figure}[hbtp]
  \setlength{\abovecaptionskip}{-0.2cm} 
  \setlength{\belowcaptionskip}{-0.2cm} 
  \centering
  \includegraphics[width = 0.45\textwidth]{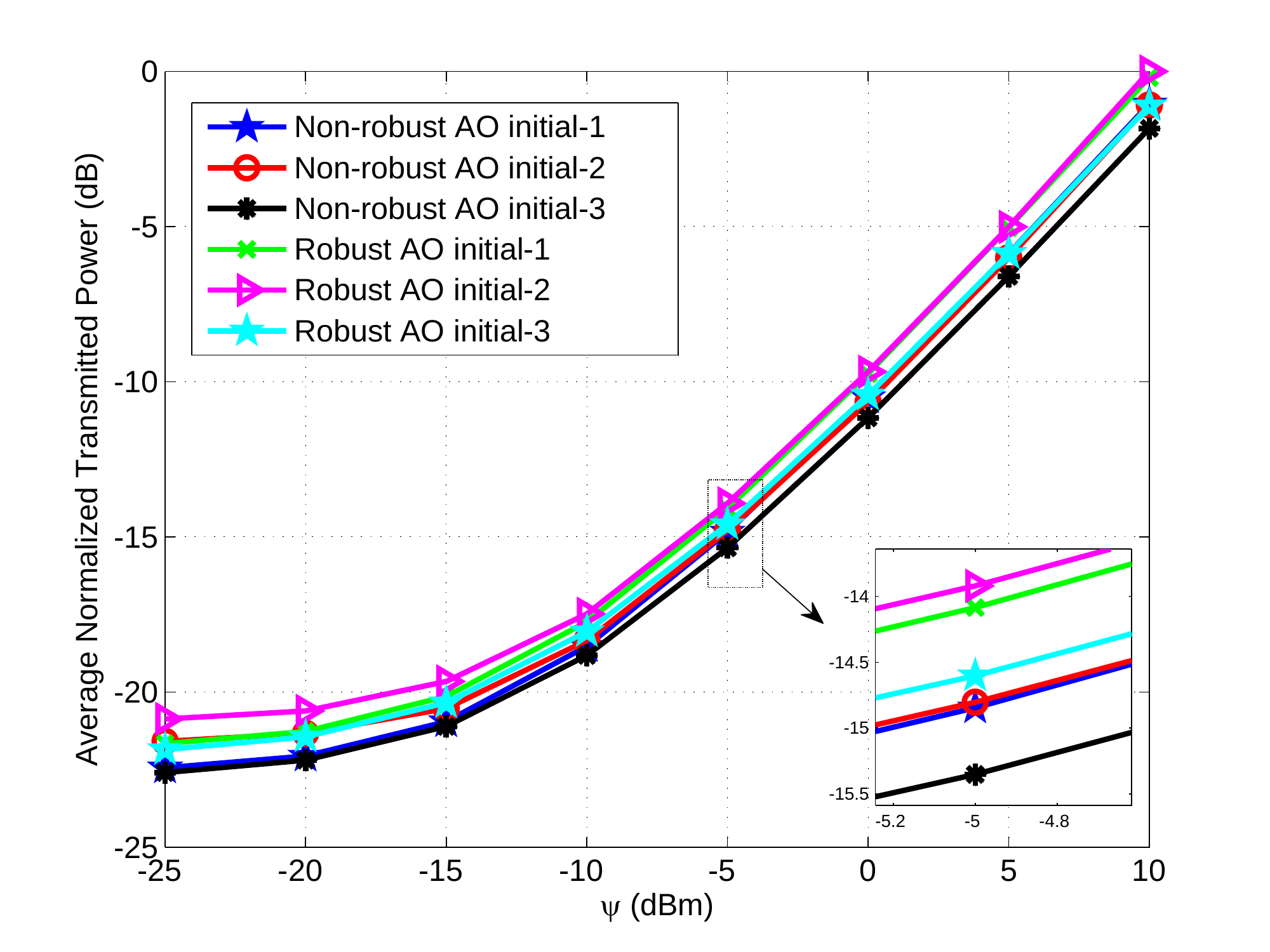}\\
  \caption{Initialization methods comparison of the AO-based transceiver designs ($\eta  = 0.1$, $\gamma = 10$dB).}\label{compare_initialization}
\end{figure}

In the AO-based transceiver design algorithms, the RS AF matrices $\bf{W}$ can be initialized in three different ways:
\begin{itemize}
\item \emph{Init-1 :} Initialize the RS AF matrix as an identity matrix.

\item \emph{Init-2 :} Random matrices are generated according to normal distribution with zero mean and unit variance.

\item \emph{Init-3 :} According to Section \ref{AO} and \ref{SR}, the SR-based transceiver can be used to initialize the AO-based designs.
\end{itemize}
Fig. \ref{compare_initialization} shows the average power consumption performance versus the receiver EH target $\psi$ for the AO-based transceiver designs with the three different initialization methods.
From the results, we can see that the best performance is achieved by \emph{Init-3} for both the non-robust and robust algorithms; \emph{Init-1} and \emph{Init-2} result into similar performance. For example, the power consumption of \emph{Init-3} is $0.5$dB less than that of \emph{Init-1} and \emph{Init-2} for the non-robust case, and the power saving rises up to $1$dB for the robust case. As seen, the AO-based algorithms are sensitive to the initial point. The use of \emph{Init-3} is not realistic in practice since it has to employ both the AO-based and the SR-based algorithms. However, in the following simulations, we employ \emph{Init-3} as the initialization method for both the non-robust and robust AO algorithms, which serves as a performance bound unless otherwise stated.

\begin{figure}[hbtp]
  \setlength{\abovecaptionskip}{-0.2cm} 
  \setlength{\belowcaptionskip}{-0.2cm} 
  \centering
  \includegraphics[width = 0.45\textwidth]{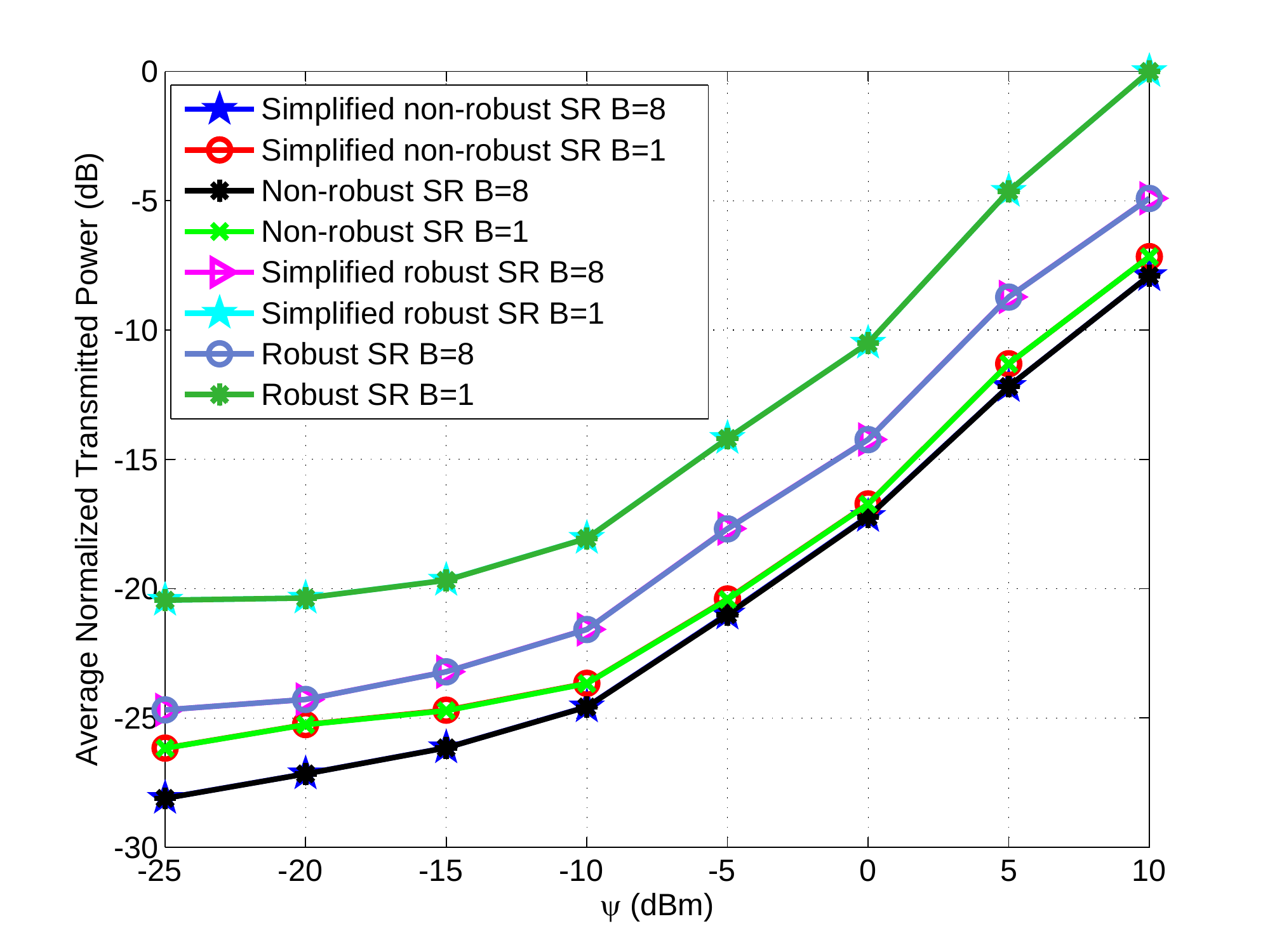}\\
  \caption{Performance comparison of the non-simplified SR-based transceiver designs and the simplified SR-based transceiver designs ($\eta  = 0.1$, $\gamma = 10$dB, ${\beta _l}(0) = 1$, $\theta = 6$).}\label{compare_simplified}
\end{figure}

Fig. \ref{compare_simplified} shows the performance comparison of the SR-based transceiver design algorithms proposed in Subsection \ref{nonrobustSR} and \ref{robustSR} and the simplified SR-based transceiver design proposed in Subsection \ref{simplified_SR}. It is worth noting that the algorithms with codebook size $B=1$ only use the identity matrix as the permutation matrix (naive method)\footnote{The naive method means that the RS only processes the received data vector ${{\bf{y}}_R}$ by adjusting its power (i.e. scaling) and forwarding it.}, and the algorithms with the $B=8$ codebook employ the \emph{Sum-Max} codebook design method. We include the identity matrix into the $B=8$ codebook regardless of the singular values in order to guarantee that the performance of the algorithms with $B=8$
is always better than with $B=1$. We can see that the simplified SR-based transceiver design algorithms achieve almost the same performance as the non-simplified ones for both non-robust and robust cases. However, the simplified SR-based algorithms consume much less computational resources compared with their non-simplified counterparts. Thus, we employ the simplified SR-based design in the following simulations for comparison unless otherwise specified.

\begin{figure}[hbtp]
  \setlength{\abovecaptionskip}{-0.2cm} 
  \setlength{\belowcaptionskip}{-0.2cm} 
  \centering
  \includegraphics[width = 0.45\textwidth]{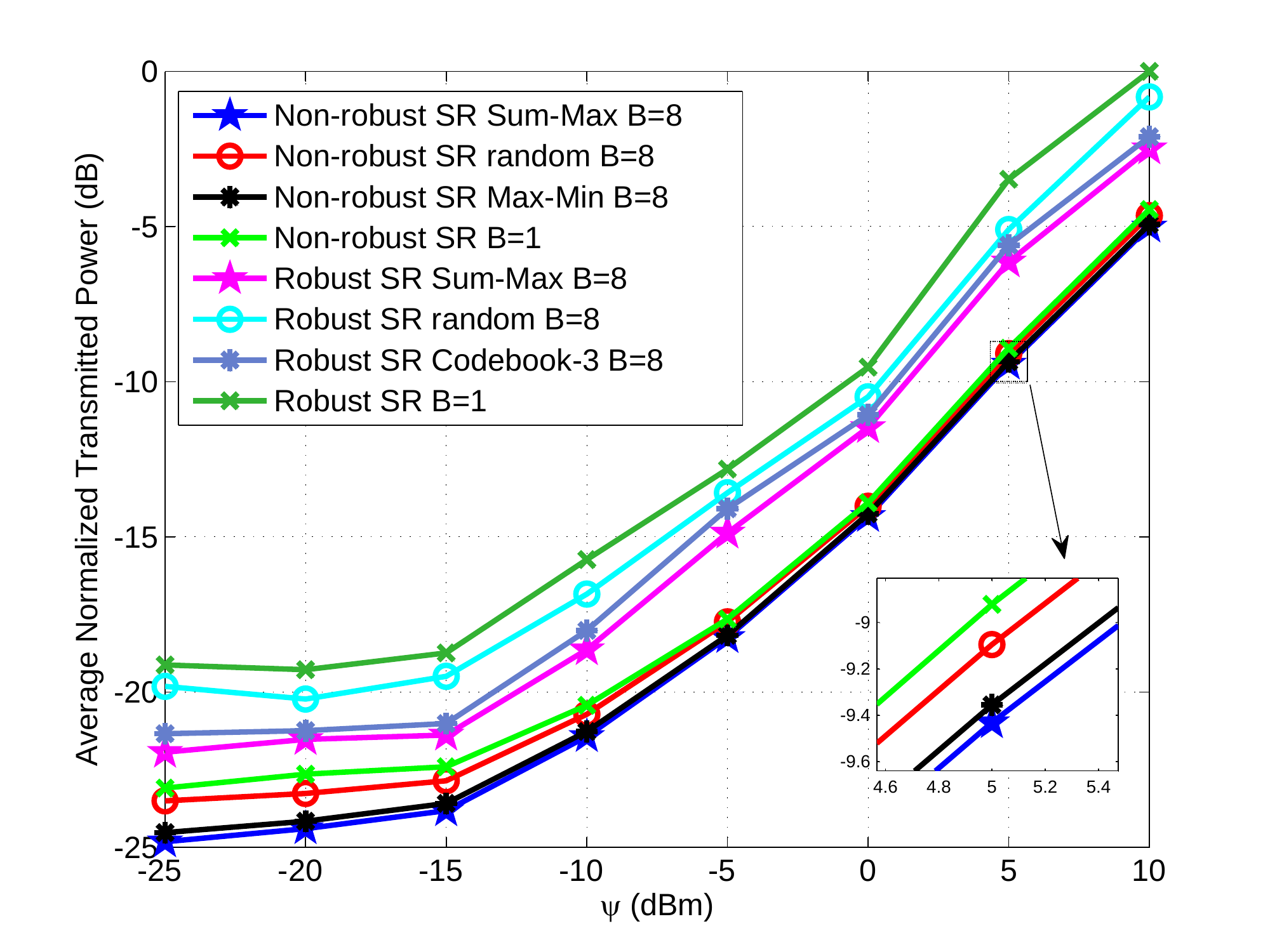}\\
  \caption{Comparison of the codebook design methods for the SR-based transceiver designs ($\eta  = 0.1$, $\gamma = 10$dB, ${\beta _l}(0) = 1$, $\theta = 6$).}\label{compare_codebookdesign}
\end{figure}

We compare the performance of the codebook design approaches, i.e., the \emph{Sum-Max} and \emph{Max-Min} methods, in terms of average power consumption. For completeness, we also consider the performance of a codebook whose elements (permutation matrices) are randomly selected.
The results which are illustrated in Fig. \ref{compare_codebookdesign} show the average power consumption performance curves versus EH target. We can see that the \emph{Sum-Max} method outperforms other methods. Compared with the randomly generated codebook, the proposed \emph{Sum-Max} method can lead to a power saving of $0.4$dB for the non-robust case and $2$dB for the robust case. The performance of the \emph{Max-Min} method is slightly inferior to the \emph{Sum-Max} method.

\begin{figure}[hbtp]
  \setlength{\abovecaptionskip}{-0.2cm} 
  \setlength{\belowcaptionskip}{-0.2cm} 
  \centering
  \includegraphics[width = 0.45\textwidth]{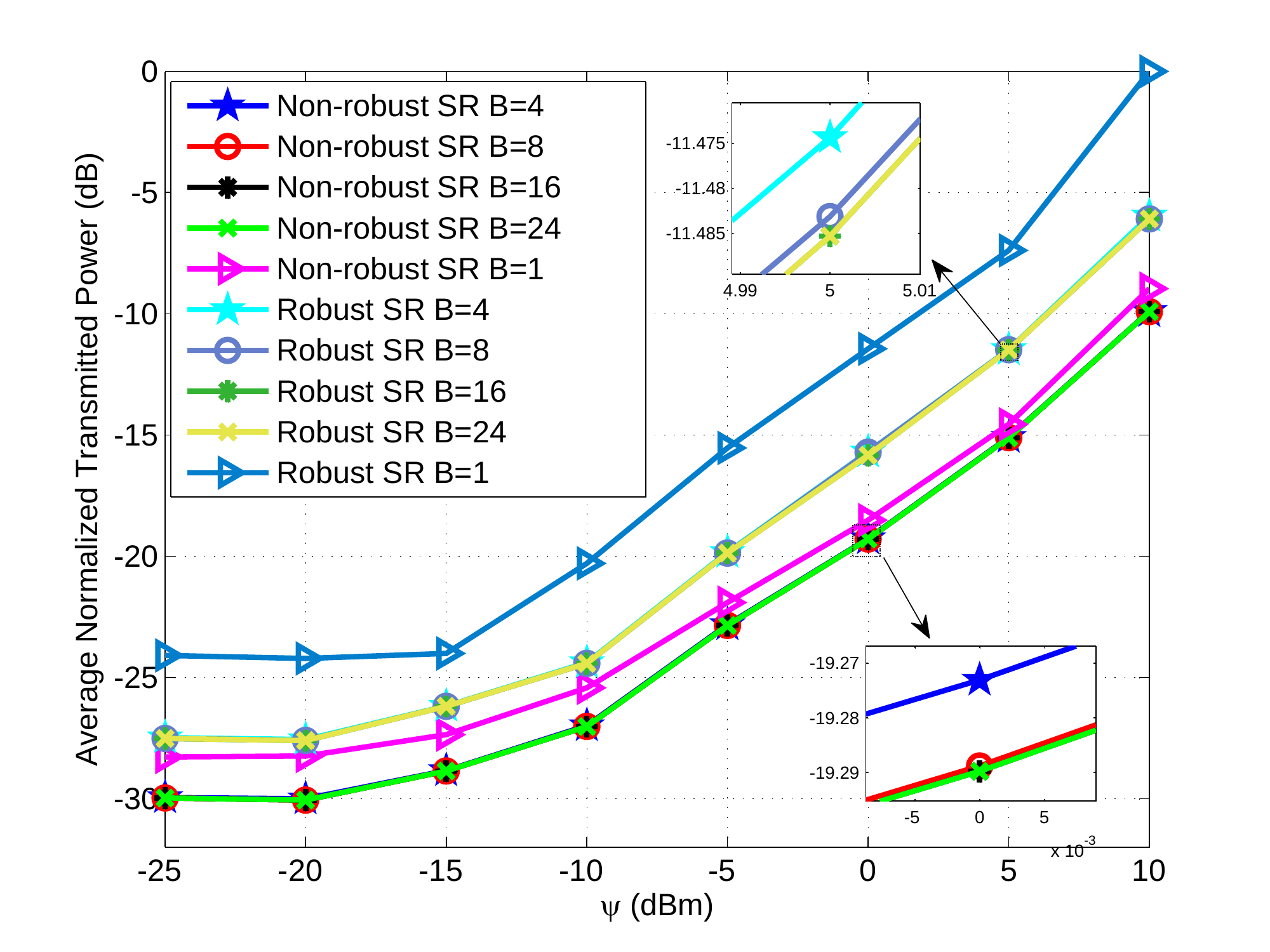}\\
  \caption{Codebook size comparison of the SR-based transceiver designs ($\eta  = 0.1$, $\gamma = 10$dB, ${\beta _l}(0) = 1$, $\theta = 6$).}\label{compare_codebooksize}
\end{figure}

Fig. \ref{compare_codebooksize} shows the average power consumption performance for various codebook sizes in the SR-based transceiver design algorithms. It is observed that the SR-based robust transceiver design algorithm with codebooks of size $B=4, 8, 16$ and $24$ achieves a power saving of $5$dB compared to that with the $B=1$ codebook, and its non-robust counterpart with codebooks of size $B=4, 8, 16$ and $24$ achieves about $1$dB in power saving compared with the $B=1$ codebook.\footnote{Note that the algorithms with $B=24$ codebook conduct a search for all the $N_r!$ permutation matrices, which is known as the optimal SR-based transceiver design algorithm.} We can also see from this figure that an increase in $B$ does not necessarily gives rise to an improvement of the power consumption performance when $B>8$, which is the reason we proposed to construct a restricted number of transceivers, i.e., to select a small fraction of all the permutation matrices for transceiver design ($B\ll N_r!$).

\begin{figure}[hbtp]
  \setlength{\abovecaptionskip}{-0.2cm} 
  \setlength{\belowcaptionskip}{-0.2cm} 
  \centering
  \includegraphics[width = 0.45\textwidth]{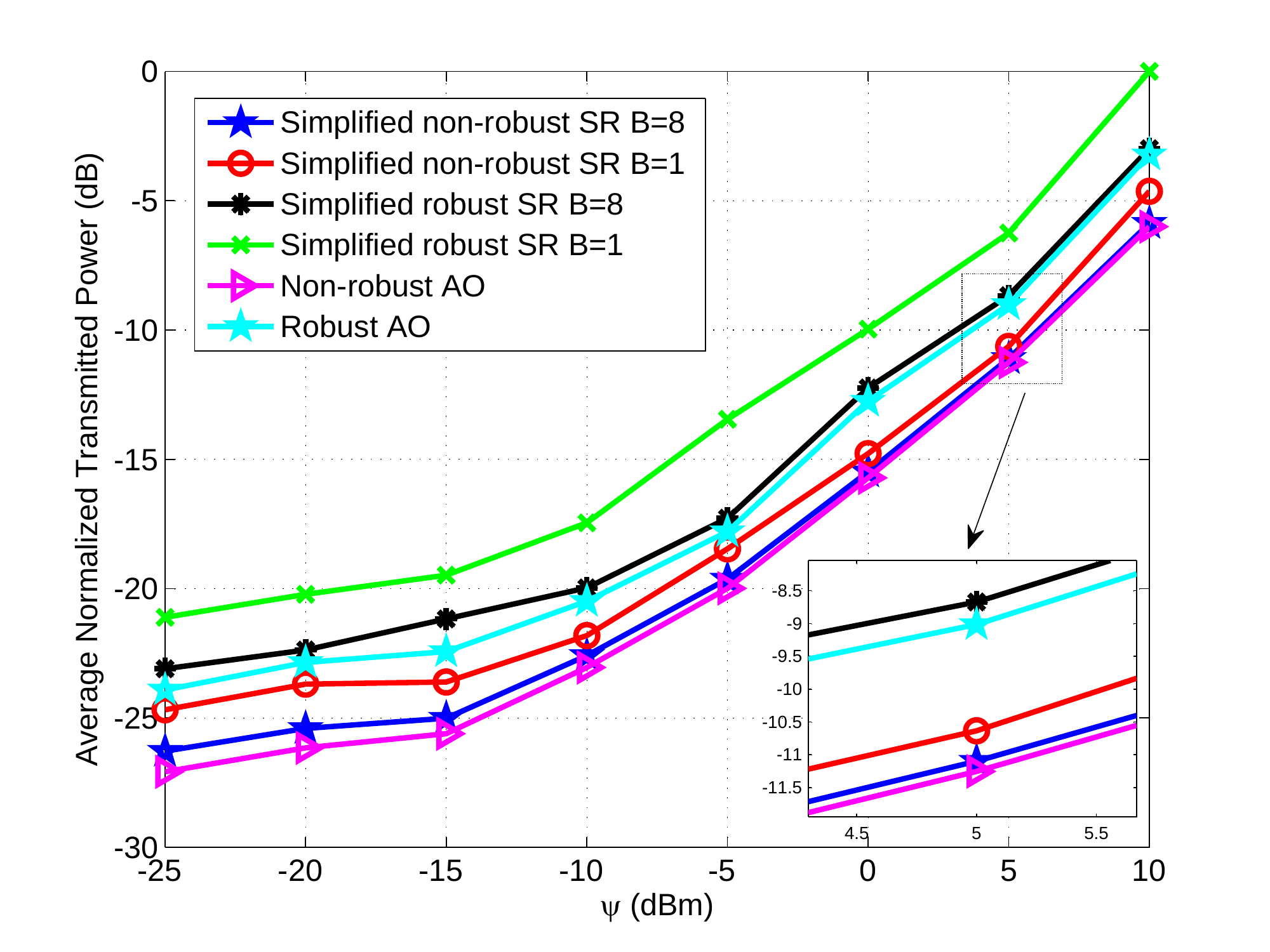}\\
  \caption{Performance comparison of the SR-based transceiver designs and the AO-based transceiver designs ($\eta  = 0.1$, $\gamma = 10$dB, ${\beta _l}(0) = 1$, $\theta = 6$).}\label{compare_algorithm}
\end{figure}

\begin{figure}[hbtp]
  \setlength{\abovecaptionskip}{-0.2cm} 
  \setlength{\belowcaptionskip}{-0.2cm} 
  \centering
  \includegraphics[width = 0.45\textwidth]{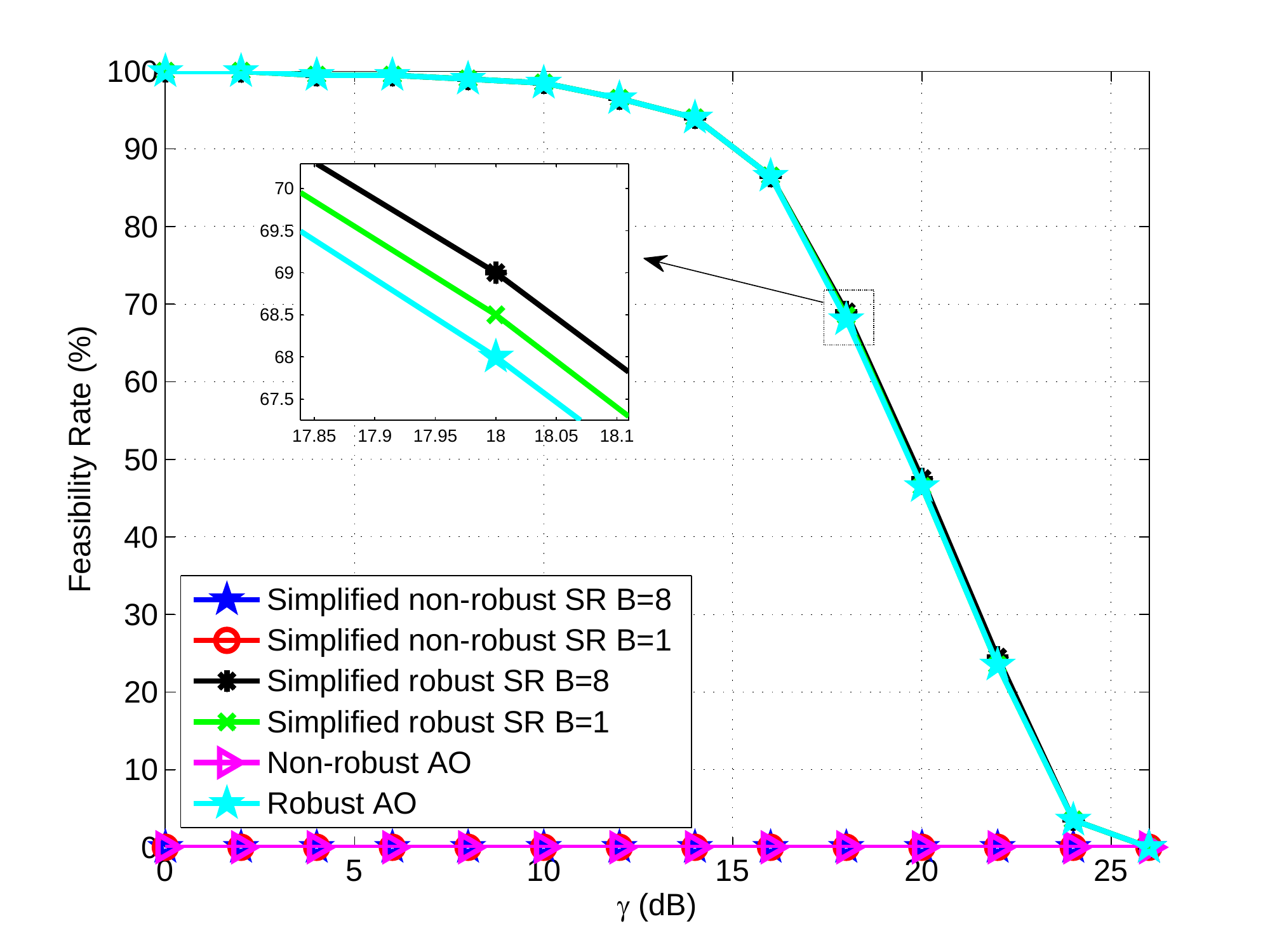}\\
  \caption{Feasibility rate comparison of the SR-based transceiver designs and the AO-based transceiver designs versus SINR threshold  $\gamma$ ($\eta  = 0.1$, ${\beta _l}(0) = 1$, $\theta = 6$).}\label{compare_fea_gamma}
\end{figure}

\begin{figure}[hbtp]
  \setlength{\abovecaptionskip}{-0.2cm} 
  \setlength{\belowcaptionskip}{-0.2cm} 
  \centering
  \includegraphics[width = 0.45\textwidth]{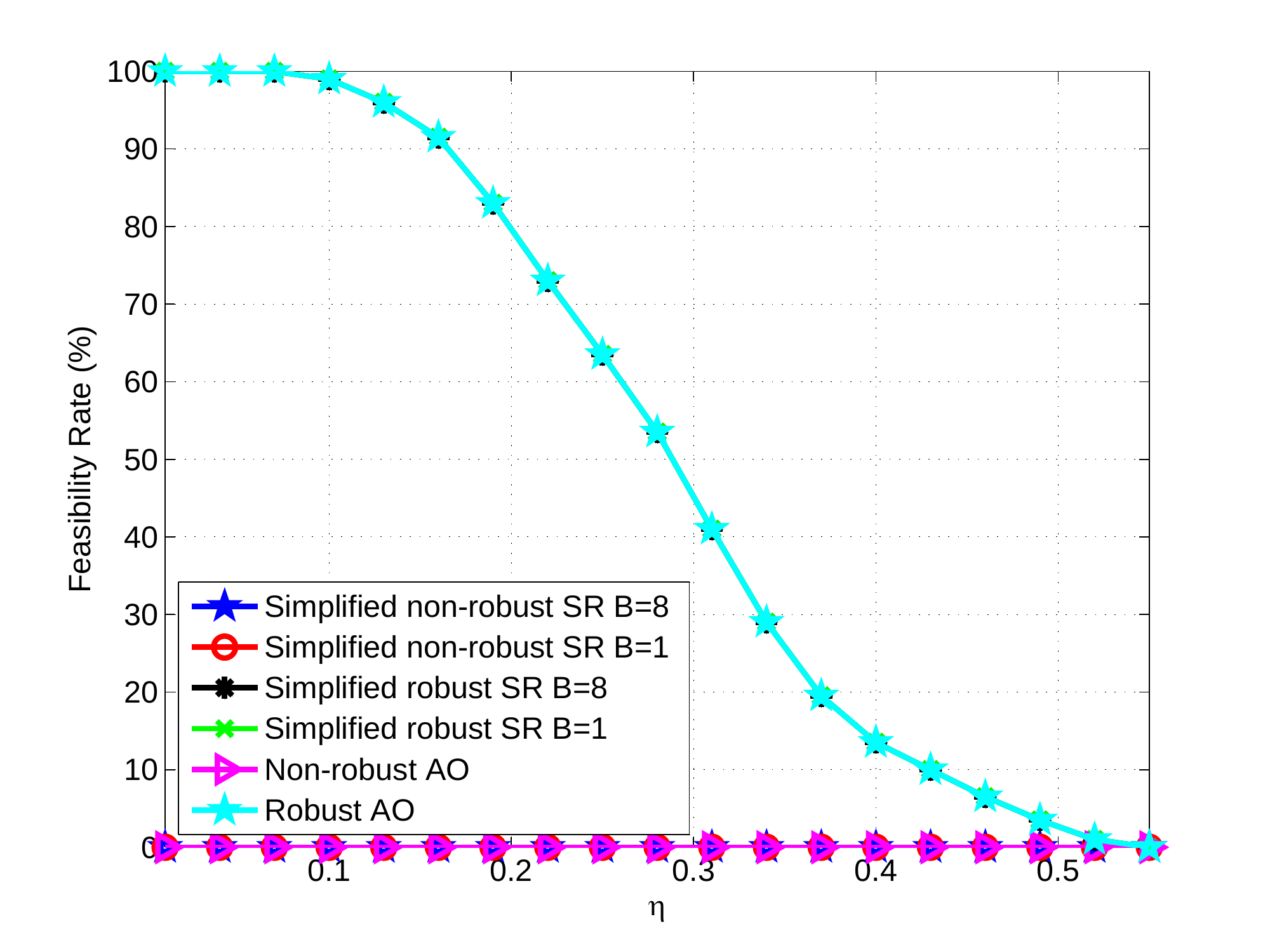}\\
  \caption{Feasibility rate comparison of the SR-based transceiver designs and the AO-based transceiver designs versus CSI error bound $\eta$ ($\gamma = 10$dB, $\psi = 5$dBm, ${\beta _l}(0) = 1$, $\theta = 6$).}\label{compare_fea_eta}
\end{figure}

In the next series of simulations, we examine the comparative performance of the AO-based and SR-based transceiver design algorithms. Fig. \ref{compare_algorithm} shows the average normalized transmitted power versus $\psi$ for these two algorithms. It is observed that the performance of the SR-based algorithms is very close to that of the AO-based algorithms.
From the complexity analysis in Section \ref{complexity}, we note that if $N_r$ increases, the complexity of solving problem (\ref{nonrobust-W}) and (\ref{robust-W}) in the AO-based transceiver design might become unacceptable, which limits the practicality of the algorithm. Thus, the SR-based transceiver design is very promising and suitable for systems with large $N_r$. It is also important to mention that the number of signaling bits of the SR-based algorithms is much less than the AO-based algorithms since only the index of the optimal transceiver and the power scaling factor have to be sent to the RS instead of the whole RS AF matrix. Besides, a comparison of the feasibility rate between the AO-based and SR-based algorithms is provided. Specifically, Fig. \ref{compare_fea_gamma} presents the feasibility rate comparison versus SINR target $\gamma$. One can observe that the feasibility rate performance of the AO-based algorithms is slightly inferior to that of the SR-based algorithms since we employ \emph{Init-3} as the initialization method for the AO-based algorithms. Fig. \ref{compare_fea_eta} presents a similar comparison of the feasibility rate versus the channel error bound $\eta$. Similarly, from this figure, we can see that the performance of the AO-based algorithms and the SR-based algorithms is very close. The non-robust algorithms fail to satisfy both the SINR and EH constraints almost all the time under the NBE model.

\section{Conclusion} \label{conclusion}
In this paper, we considered the joint transceiver design problem for multiuser MISO relay systems with energy harvesting. We proposed AO-based and SR-based algorithms, including both non-robust and robust versions to CSI errors, for the joint optimization of the BS beamforming vectors, the RS AF matrix and the receiver PS ratios. Especially, the SR-based algorithms can achieve almost the same performance compared with the AO-based algorithms but with much reduced computational complexity and overhead. We also presented a simplified SR-based algorithm and carried out a detailed complexity analysis of all the proposed algorithms.
Two efficient approaches for the design of the permutation matrix codebook in the SR-based algorithms were proposed.
The simulation results validated the effectiveness of the proposed joint transceiver design algorithms and the necessity of using a robust formalism in the presence of imperfect CSI. We note that the proposed algorithms can be easily extended to more complicated (multi-hop) relay systems, which remains an open avenue for future work.
\begin{appendices}
\section{Proposed Rank-one Recovery Method} \label{rankonerecovery}
As we mentioned in Section \ref{AO} and \ref{SR}, the optimal solution of problem (\ref{nonrobust-W}), (\ref{robust-Fk}), (\ref{robust-W}) and (\ref{robust_fixedbeta}) is not guaranteed to be rank-one. In this appendix, we propose a rank-one recovery method based on a randomization procedure. Inspired by the concept of worst-case robustness {\cite{li2003robust, li2004doubly, kim2008robust}}, we consider the following subproblems:
\begin{equation} \label{worst-case1}
{u_{kj}} = \mathop {\min }\limits_{\|{{\bf{e}}_k}\|^2 \le \eta _k^2} \;{| {(\widehat {\bf{h}}_k^H + {\bf{e}}_k^H){{\bf{W}}^*}{\bf{Gf}}_j^*} |^2},
\end{equation}
\begin{equation}\label{worst-case2}
{v_{kj}} = \mathop {\max }\limits_{\|{{\bf{e}}_k}\|^2 \le \eta _k^2} \;{| {(\widehat {\bf{h}}_k^H + {\bf{e}}_k^H){{\bf{W}}^*}{\bf{Gf}}_j^*} |^2},\;j \ne k,
\end{equation}
\begin{equation}\label{worst-case3}
{\overline w_k } = \mathop {\max }\limits_{\|{{\bf{e}}_k}\|^2 \le \eta _k^2} \;{\| {(\widehat {\bf{h}}_k^H + {\bf{e}}_k^H){{\bf{W}}^*}} \|^2},
\end{equation}
\begin{equation}\label{worst-case4}
{\widetilde w_k} = \mathop {\min }\limits_{\|{{\bf{e}}_k}\|^2 \le \eta _k^2} \;{\| {(\widehat {\bf{h}}_k^H + {\bf{e}}_k^H){{\bf{W}}^*}} \|^2},
\end{equation}
where ${{\bf{f}}_k^*}$ and ${{\bf{W}}^*}$ are extracted from the optimal solutions ${{\bf{F}}_k^*}$ and ${\widetilde {\bf{W}}^*}$ using the randomization procedure as will be detailed in Table \ref{rank-one-recovery}.\footnote{For problem (\ref{robust_fixedbeta}), ${{\bf{W}}^*}$ can be expressed as $\sqrt{\beta _l^*}{{\bf{T}}_l}$, where $\beta _l^*$ denotes the relay power scaling factor corresponding to the $l$th latent transceiver.} It is not difficult to see that problem (\ref{worst-case1})-(\ref{worst-case4}) can be solved by the Cauchy-Schwarz inequality and Lagrange multiplier method. We omit the detailed derivation and let $\{ u_{kj}^*,v_{kj}^*,{\overline w_k^*},{\widetilde w_k^*}\} $ denote the corresponding optimal objective values.

For problem (\ref{robust-Fk}) and (\ref{robust_fixedbeta}), we propose to scale up ${\bf{f}}_k^*$ by $\sqrt{\varphi_k}$ and then jointly optimize $\{\sqrt{\varphi_k}\}$ and $\{\rho_k\}$ to recover a rank-one solution. The optimization problem can be formulated as
\begin{equation}
\begin{array}{l}
\mathop {\min }\limits_{\{ {\varphi _k},\;{\rho _k}\} } \;\sum\limits_{k = 1}^K {{\varphi _k}{\bf{f}}_k^{*H}{\bf{f}}_k^*}  + \sum\limits_{k = 1}^K {{\varphi _k}{{\left\| {{\bf{W}^*\bf{Gf}}_k^*} \right\|}^2}} \\
\textrm{s.t.}\;\frac{{{\rho _k}{\varphi _k}u_{kk}^*}}{{{\rho _k}\sum\limits_{j \ne k}^K {{\varphi _j}v_{kj}^*}  + {\rho _k}\sigma _r^2{\overline w_k^*} + {\rho _k}\sigma _k^2 + \omega _k^2}} \ge {\gamma _k},\\
\sum\limits_{j = 1}^K {{\varphi _j}u_{kj}^*}  + \sigma _r^2\widetilde w_k^* + \sigma _k^2 \ge \frac{{{\psi _k}}}{{{\xi _k}(1 - {\rho _k})}},\\
{\varphi _k} \ge 0,\;0 \le {\rho _k} \le 1,\;\forall k \in \mathcal{K}.
\end{array}
\end{equation}
The above problem can be reformulated as the following SOCP problem
\begin{equation} \label{rank-one-recovery-1}
\begin{array}{l}
\mathop {\min }\limits_{\{ {\varphi _k},\;{\rho _k}\} } \;\sum\limits_{k = 1}^K {{\varphi _k}{\bf{f}}_k^{*H}{\bf{f}}_k^*}  + \sum\limits_{k = 1}^K {{\varphi _k}{{\| {{\bf{W}^*\bf{Gf}}_k^*} \|}^2}} \\
\textrm{s.t.}\;\| {{{[2{\omega _k},{z_k} - {\rho _k}]}}} \| \le {z_k} + {\rho _k},\\
\| {{{[2\sqrt {{\psi _k} / {\xi _k}} ,{{\widetilde z}_k} - 1 + {\rho _k}]}}} \| \le {\widetilde z_k} + 1 - {\rho _k},\\
{z_k} = \frac{{{\varphi _k}u_{kk}^*}}{{{\gamma _k}}} - \sum\limits_{j \ne k}^K {{\varphi _j}v_{kj}^*}  - \sigma _k^2 - \sigma _r^2{\overline w_k^*},\\
{\widetilde z_k} = \sum\limits_{j = 1}^K {{\varphi _j}u_{kj}^*}  + \sigma _r^2\widetilde w_k^* + \sigma _k^2,\\
{\varphi _k} \ge 0,\;0 \le {\rho _k} \le 1,\;\forall k \in \mathcal{K}.
\end{array}
\end{equation}

Similarly, we also propose to scale up ${\bf{W}}^*$ by $\sqrt{\varphi}$ and then jointly optimize $\{\sqrt{\varphi}\}$ and $\{\rho_k\}$ to recover a rank-one solution for problem (\ref{nonrobust-W}) and (\ref{robust-W}). The optimization problem can be formulated as\footnote{For problem (\ref{nonrobust-W}), $\{ u_{kj}^*,v_{kj}^*,{\overline w_k^*},{\widetilde w_k}^*\} $ is the solution of (\ref{worst-case1})-(\ref{worst-case4}) with ${\bf{e}}_k={\bf{0}},\; \forall k$.}
\begin{equation} \label{rank-one-recovery-2}
\begin{array}{l}
\mathop {\min }\limits_{\{ \varphi ,{\rho _k}\} } \; \sum\limits_{k = 1}^K {\varphi {{\| {{{\bf{W}}^*}{\bf{Gf}}_k^*} \|}^2}}  + \sigma _r^2\varphi {\| {{{\bf{W}}^*}} \|^2}\\
\textrm{s.t.}\;\frac{{{\rho _k}\varphi u_{kk}^*}}{{{\rho _k}\varphi \sum\limits_{j \ne k}^K {v_{kj}^*}  + \varphi {\rho _k}\sigma _r^2{\overline w_k^*} + {\rho _k}\sigma _k^2 + \omega _k^2}} \ge {\gamma _k},\\
\varphi \sum\limits_{j = 1}^K {u_{kj}^*}  + \varphi \sigma _r^2\widetilde w_k^* + \sigma _k^2 \ge \frac{{{\psi _k}}}{{{\xi _k}(1 - {\rho _k})}},\\
\varphi  \ge 0,\;0 \le {\rho _k} \le 1,\;\forall k \in \mathcal{K}.
\end{array}
\end{equation}
As has been discussed in our previous work \cite{Mingmin2015robust}, the above optimization problem admits a closed-form solution and we omit the detailed derivation in this work. The proposed rank-one recovery method is summarized in Table \ref{rank-one-recovery}.\footnote{We only list the detailed steps to recover ${\bf{f}}_k^*$, which can be easily extended to the recovery of ${{{\bf{W}}^*}}$.}
\begin{table}
  \centering
  \caption{Proposed rank-one recovery method} \label{rank-one-recovery}
  \begin{tabular}{p{0.9\columnwidth}}
 \hline
 \begin{itemize}
 \item[1.] \textbf{For} the $i$th randomization step ($i=1,\ldots,R$, $R$ is the number of randomization steps.)
 \item[2.] \begin{itemize}
 \item If $i=1$, let ${\bf{f}}_k^*$ equal to the principal component of ${\bf{F}}_k^*$. Else, calculate the eigenvalue decomposition ${\bf{F}}_k^* = {\bf{Q\Lambda }}{{\bf{Q}}^H}$, then the candidate vector is generated as ${\bf{f}}_k^* = {\bf{Q}}{{\bf{\Lambda }}^{\frac{1}{2}}}{\bf{w}}$, where ${\bf{w}}$ is an i.i.d. complex Gaussian vector with zero mean and unit variance.
 \item  Solve problem (\ref{rank-one-recovery-1}) to obtain the optimal $\{\varphi _k^*, \rho_k^*\}$. If the problem turns out to be infeasible, discard that candidate vector. Otherwise, save the candidate $\{{\bf{f}}_k^*,\varphi _k^*, \rho_k^*\}$ and the objective value of the problem.
 \end{itemize}
 {\textbf{End}}
 \item[4.] Select the candidate $\{{\bf{f}}_k^*,\varphi _k^*, \rho_k^*\}$ which corresponds to the minimum objective value.
 \end{itemize} \\
  \hline
 \end{tabular}
\end{table}

\section{The proof of Proposition 1} \label{appendixB}
We first introduce four sets of auxiliary variables ${d_k}$, $\widetilde d_k$, ${e_k}$ and $\widetilde e_k$, $\forall k \in \mathcal{K}$, which satisfy
\begin{equation}
{d_k} = \sigma _r^2{\bf{h}}_k^H{{\bf{h}}_k} + \frac{1}{4}\omega _k^2{(p_k^{(i)} - \varphi _l^{(i)})^2} + \frac{1}{{{\gamma _k}}}{\bf{f}}_k^{(i)H}{\widetilde {\bf{G}}_k}{\bf{f}}_k^{(i)},
\end{equation}
\begin{equation}
\widetilde d_k =  - \sigma _k^2{\varphi _l} + \frac{1}{2}(p_k^{(i)} - \varphi _l^{(i)})(p_k^{} - \varphi _l^{}) + \frac{2}{{{\gamma _k}}}\mathcal{R}\{ {\bf{f}}_k^{(i)H}{\widetilde {\bf{G}}_k}{\bf{f}}_k^{}\},
\end{equation}
\begin{equation}
{e_k} =  - \sigma _r^2{\bf{h}}_k^H{{\bf{h}}_k} + \sum\limits_{j = 1}^K {{\bf{f}}_j^{(i)H}{{\widetilde {\bf{G}}}_k}{\bf{f}}_j^{(i)}}  + \frac{{{\psi _k}}}{{4{\xi _k}}}{(q_k^{(i)} - \varphi _l^{(i)})^2},
\end{equation}
\begin{equation}
\widetilde e_k = 2{\mathop{\mathcal{R}}\nolimits} \{ \sum\limits_{j = 1}^K {{\bf{f}}_j^{(i)H}{{\widetilde {\bf{G}}}_k}{{\bf{f}}_j}} \}  + \frac{{{\psi _k}}}{{2{\xi _k}}}(q_k^{(i)} - \varphi _l^{(i)})(q_k^{} - \varphi _l^{}) + \sigma _k^2{\varphi _l},
\end{equation}
where ${d_k}$ and ${e_k}$ are constant scalars while $\widetilde d_k$ and $\widetilde e_k$ are affine functions in ${\bf{r}}$. Then, the SINR constraints in problem (\ref{nonrobust-CCCP1}) can be rewritten as the following SOCs
\begin{equation} \label{sinr-socp}
\left\|\left[{\bf{w}}_k^T,\frac{{\widetilde d_k - {d_k} - 1}}{2}\right]\right\| \le \frac{{\widetilde d_k - {d_k} + 1}}{2},\forall k,
\end{equation}
where
\begin{equation} \label{definitionw}
\begin{array}{*{20}{l}}
{{{\bf{w}}_k} = [{\bf{h}}_k^H{{\bf{T}}_l}{\bf{G}}{{\bf{f}}_1}, \ldots, {\bf{h}}_k^H{{\bf{T}}_l}{\bf{G}}{{\bf{f}}_{k - 1}},{\bf{h}}_k^H{{\bf{T}}_l}{\bf{G}}{{\bf{f}}_{k + 1}},}\\
{ \ldots ,{\bf{h}}_k^H{{\bf{T}}_l}{\bf{G}}{{\bf{f}}_K},\frac{1}{2}{\omega _k}({p_k} + {\varphi _l}){]^T}}.
\end{array}
\end{equation}
Similarly, the EH constraints in problem (\ref{nonrobust-CCCP1}) can be reformulated as the following SOCs
\begin{equation} \label{eh-socp}
\left\|\left[\sqrt {\frac{{{\psi _k}}}{{4{\xi _k}}}} ({q_k} + {\varphi _l}),\frac{{\widetilde e_k - {e_k} - 1}}{2}\right]\right\| \le \frac{{\widetilde e_k - {e_k} + 1}}{2},\forall k.
\end{equation}

Next, by introducing slack variables $P_1$, $P_2$ and $P_3$, we can equivalently write the objective function $P_l({\bf{r}})$ of problem (\ref{nonrobust-CCCP1}) as follows
\begin{equation}
P_l({\bf{r}}) = {P_1} + {P_2} + {P_3},
\end{equation}
where
\begin{equation} \label{power1}
\| {[{\bf{f}}_1^T, \ldots ,{\bf{f}}_K^T,({P_1} - 1)/2]} \| \le ({P_1} + 1)/2,
\end{equation}
\begin{equation} \label{power2}
\| {[{{({\bf{G}}{{\bf{f}}_1})}^T}, \ldots ,{{({\bf{G}}{{\bf{f}}_K})}^T},({P_2} - {\varphi _l})/2]} \| \le ({P_2} + {\varphi _l})/2,
\end{equation}
\begin{equation} \label{power3}
\| {[2{\sigma _r},{\varphi _l} - {P_3}]} \| \le {\varphi _l} + {P_3}.
\end{equation}
Therefore, problem (\ref{nonrobust-CCCP1}) can be shown to be equivalent to the following SOCP problem
\begin{equation}
\begin{array}{*{20}{l}}
{\mathop {\min }\limits_{{\bf{r}},\;{P_1},\;{P_2},\;{P_3}} \;{P_1} + {P_2} + {P_3}}\\
{{\textrm{s}}.{\textrm{t}}.\;(\ref{sinr-socp}),\; (\ref{eh-socp}),\;(\ref{power1}),\;(\ref{power2})\; \textrm{and}\; (\ref{power3}) },\\
{{p_k} \ge 1,\;{q_k} \ge 1,\;{\textrm{invp}}({p_k}) + {\textrm{invp}}({q_k}) \le 1,}\;\forall k \in \mathcal{K}.\\
\end{array}
\end{equation}
This completes the proof.

\end{appendices}

\bibliography{references}

\end{document}